\documentclass[aps,prl,superscriptaddress,amsmath,amssymb,aps,twocolumn,]{revtex4-2}
\usepackage{graphicx}
\usepackage{dcolumn}
\usepackage{bm}
\usepackage{latexsym,epsfig}
\usepackage{graphicx}
\usepackage{verbatim}
\usepackage{comment}
\usepackage{amsmath}
\usepackage{amsfonts}
\usepackage{amsthm}
\usepackage{amssymb}
\usepackage{mathrsfs}
\usepackage{stmaryrd}
\usepackage{color}
\usepackage{physics}
\usepackage{epstopdf}
\usepackage{grffile}
\usepackage{bbold}
\usepackage[normalem]{ulem}
\usepackage{mathtools}
\DeclareGraphicsExtensions{.eps}
\newcommand{\beq}{\begin{equation}}
\newcommand{\eeq}{\end{equation}}
\newcommand{\bea}{\begin{eqnarray}}
\newcommand{\eea}{\end{eqnarray}}
\newcommand{\ben}{\begin{eqnarray*}}
\newcommand{\een}{\end{eqnarray*}}
\newcommand{\bfig}{\begin{figure}}
\newcommand{\efig}{\end{figure}}

\newcommand{\ra}{\rangle}
\newcommand{\la}{\langle}

\newcommand{\non}{\nonumber}

\newcommand{\de}{\delta}

\newcommand{\si}{\sigma}

\newcommand{\da}{\dagger}

\usepackage{hyperref}
\hypersetup{
 colorlinks=true, 
 urlcolor=blue,
 citecolor=blue,
 linkcolor=blue
}

\begin{document}
\title{Statistics-controlled topological phases in a Su-Schrieffer-Heeger ladder }
\author{Rajashri Parida}
\email{rajashriparida33@gmail.com}
\affiliation{School of Physical Sciences, National Institute of Science Education and Research, Jatni 752050, India}
\affiliation{Homi Bhabha National Institute, Training School Complex, Anushaktinagar, Mumbai 400094, India}
\author{Biswajit Paul}
\email{biswajitt.bp@gmail.com}
\affiliation{School of Physical Sciences, National Institute of Science Education and Research, Jatni 752050, India}
\affiliation{Homi Bhabha National Institute, Training School Complex, Anushaktinagar, Mumbai 400094, India}
\author{Smitha Vishveshwara}
\email{smivish@illinois.edu}
\affiliation{Department of Physics, University of Illinois at Urbana-Champaign, Urbana, Illinois 61801-3080, USA}
\author{Diptiman Sen}
\email{sen.diptiman@gmail.com}
\affiliation{Centre for High Energy Physics, Indian Institute of Science, Bengaluru 560012, India}
\author{Tapan Mishra}
\email{mishratapan@gmail.com}
\affiliation{School of Physical Sciences, National Institute of Science Education and Research, Jatni 752050, India}
\affiliation{Homi Bhabha National Institute, Training School Complex, Anushaktinagar, Mumbai 400094, India}

\date{\today}

\begin{abstract}

In strictly one-dimensional lattices, free fermions and hard-core bosons exhibit equivalent physical properties in many-particle systems. However, this statistical equivalence generally breaks down upon moving to higher dimensions. Our work investigates the significant effects of this difference on the topological phase transition in a minimal system lying away from one dimension: the two-leg Su-Schrieffer-Heeger (SSH) ladder. Starting in the limit of two decoupled SSH chains having opposite hopping dimerization patterns (staggered dimerization), we demonstrate that introducing inter-leg coupling drives a transition from a topological to a trivial phase for both spinless fermions and hard-core bosons. However, the topological to trivial phase transition boundary is strongly sensitive to the underlying particle statistics. We find that this intriguing effect favors a trivial phase sandwiched between the two topological phases with opposite characters for the case of hard-core bosons, resulting in a reentrant topological phase transition. In contrast, this reentrant phenomenon is absent for spinless fermions, with the topological phases being separated by a gapless line instead of an intervening trivial phase. This study demonstrates the interplay between particle statistics and topological features, revealing fundamentally different properties in an experimentally accessible set-up.

\end{abstract}
\maketitle
\paragraph*{Introduction.-} 
\label{sec:intro}

Understanding the behavior of many body quantum systems often involves comparing particles governed by fundamentally different quantum statistics, yet confined by similar physical constraints. Spinless fermions and hard-core bosons on a lattice represent such a comparative study where, despite sharing an identical exclusion constraint -- no more than one particle per lattice site -- the underlying symmetries lead to markedly different physical properties~\cite{Rigol2005,cheong_henley}. While spinless fermions naturally satisfy the Pauli exclusion principle, hard-core bosons are bosonic particles subjected to an infinite on-site repulsive interaction that prevents double occupancy.

In one dimension (1D), hard-core bosons act as if they are fermions because they cannot pass through each other~\cite{Girardeau_jmp}. This feature allows them to be mapped onto non-interacting spinless fermions via the Jordan-Wigner transformation~\cite{Jordan1928}. Despite this mapping, the fundamental difference in particle statistics (antisymmetric versus symmetric wavefunctions) means that while their energy spectra and local densities are often identical in 1D, their non-local properties differ significantly. However, in higher dimensions (e.g., 2D or 3D), this equivalence breaks down, resulting in distinctly different quantum phases for the two kinds of quantum particles.

\begin{figure}[t]
\begin{center}
\includegraphics[width=1.\columnwidth]{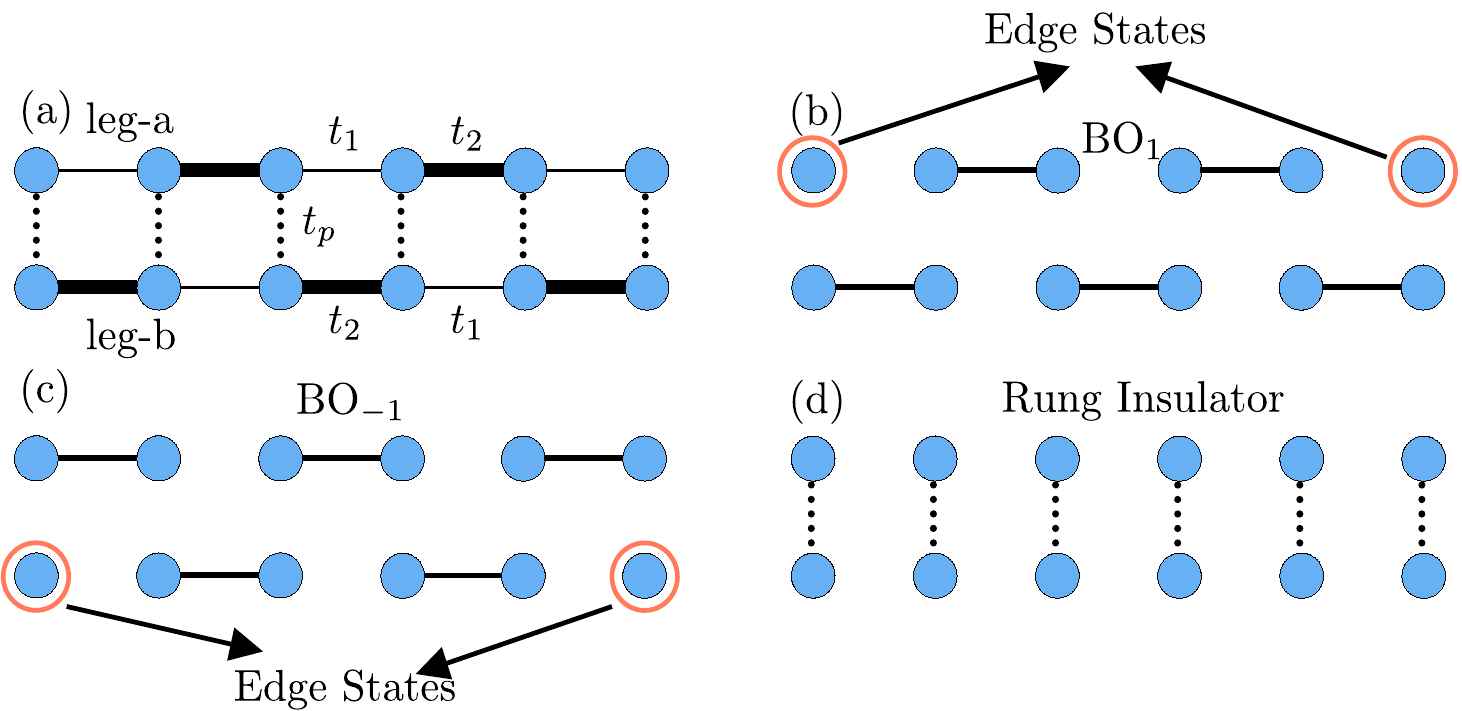}
\end{center}
\caption{(a) Schematic of the two-leg SSH ladder. (b) Topological bond order phase with edge modes localized on the upper leg  (BO$_1$). (c) Topological bond order phase with edge modes localized on the lower leg (BO$_{-1}$). (d) Schematic of the rung insulator phase. The edge states reside on the upper (lower) leg in the BO$_{1}$ (BO$_{-1}$) phase.  For both fermions and bosons, the rung insulator is characterized in terms of a single particle being dimerized on a single rung, and it can not move through the entire ladder except the rung in which it is trapped.}
\label{fig:cartoon}
\end{figure}

In this work we examine the build-up of this fundamental difference elicited by quantum statistics as a function of dimensionality for the case of
symmetry-protected topological phase transitions in the framework of the SSH model, which is one of the simplest examples exhibiting a topological phase transition in one dimension~\cite{ssh_model,Asboth2016_ssh,Atala2013,Takahashi2016pumping,Lohse2016,Mukherjee2017,Lu2014,ssh_expt_1,ssh_expt_2,browyes,EC1, EC2}. While the topological properties of the one-dimensional SSH model at half filling are identical for hard-core bosons and spinless fermions, we show that for a two-leg SSH ladder, which constitutes an immediate deviation from one dimension, the two statistically different constituents exhibit completely different topological phase transitions in their ground states. Starting from a topological ladder, we show that spinless fermions undergo a phase transition from one topological (BO$_{-1}$ in Fig.~\ref{fig:cartoon}(c)) to another distinct topological phase (BO$_{1}$ in Fig.~\ref{fig:cartoon}(b)) through a gapless critical point for a range of values of the inter-leg coupling strength. In contrast, when hard-core bosons are considered, the gapless critical point is replaced by a trivial gapped phase (Rung insulator phase in Fig.~\ref{fig:cartoon}(d)), resulting in a topological-to-trivial and a subsequent trivial-to-topological phase transition, dubbed a reentrant topological phase transition. The two topological phases, BO$_{-1}$ and BO$_1$, are distinguished by opposite signs of the topological invariant and correspond to bond-ordered states with alternating patterns of nearest-neighbor bond strengths [See Fig.~\ref{fig:cartoon}(b) and (c)]. In contrast, the rung-insulating phase is characterized by the preferential dimerization of a particle on a single rung, as shown in  Fig.~\ref{fig:cartoon}(d).
This fundamental difference in the ground state properties due to the difference in particle statistics of a many-body system is our central result. For both cases, beyond a critical inter-leg coupling, one of the topological phases disappears, and the system exhibits a topological to trivial phase transition In what follows, we introduce the model, present the contrast phase diagrams for the fermion and boson phases, analyze the diagrams in terms of energy gap closures, and via two different measures, characterize the topological nature of the phases.  

\paragraph*{Model and Approach.-}
\label{sec:model}

The Hamiltonian for the staggered SSH  ladder system is given by
\begin{align}
    H &= H_{||}-\bigg(t_p\sum_i a_i^\dagger b_i+\text{H.c.}\bigg),
    \label{eq:ham}
\end{align}
\begin{align}
    H_{||} &= -\bigg{(}t_1 \sum_{i\in\text{odd}} a_i^\dagger a_{i+1}+ t_2 \sum_{i\in\text{even}} a_i^\dagger a_{i+1} \nonumber \\
    & ~~~~~~~+t_2\sum_{i\in\text{odd}} b_i^\dagger b_{i+1} +t_1\sum_{i\in\text{even}} b_i^\dagger b_{i+1}\bigg{)}+\text{H.c.},
    \label{eq:ham_leg}
\end{align}
with $i$ representing the rung index of the ladder. 
Here $a^\dagger_{i}$ and $b_{i}$ are the particle (bosonic or fermionic) creation and annihilation operators at the $i$-th site of leg-a and leg-b, respectively, as shown in Fig.~\ref{fig:cartoon}(a). 
The hopping strength on the rungs of the ladder is $t_p$, while $t_1$ and $t_2$ denote the alternating hopping strengths along each leg.
The hopping dimerization strength is determined by the ratio between $t_1$ and $t_2$.
The ladder is dimerized for $t_1\neq t_2$, whereas $t_1=t_2$ corresponds to the absence of dimerization.
We set $t_1=1$, which defines the energy scale of the system.

We study the ground state properties using DMRG~\cite{white1992,schollowck_dmrg_rev,Verstraete_rev,schollowck_mps} with open boundary condition
(up to $L=140$ rungs or total $280$ sites, bond dimension $700$), and exact diagonalization (ED) with periodic boundary condition for $L=10$ rungs) for hard-core bosons. 
Unless noted, all the results are extrapolated to the thermodynamic limit ($L \to \infty$) to reduce finite-size effects.
We also complement our numerical results with analytical insights in certain limits.
For spinless fermions, all the calculations are exact.

\begin{figure}[t]
\begin{center}
\includegraphics[width=1.\columnwidth]{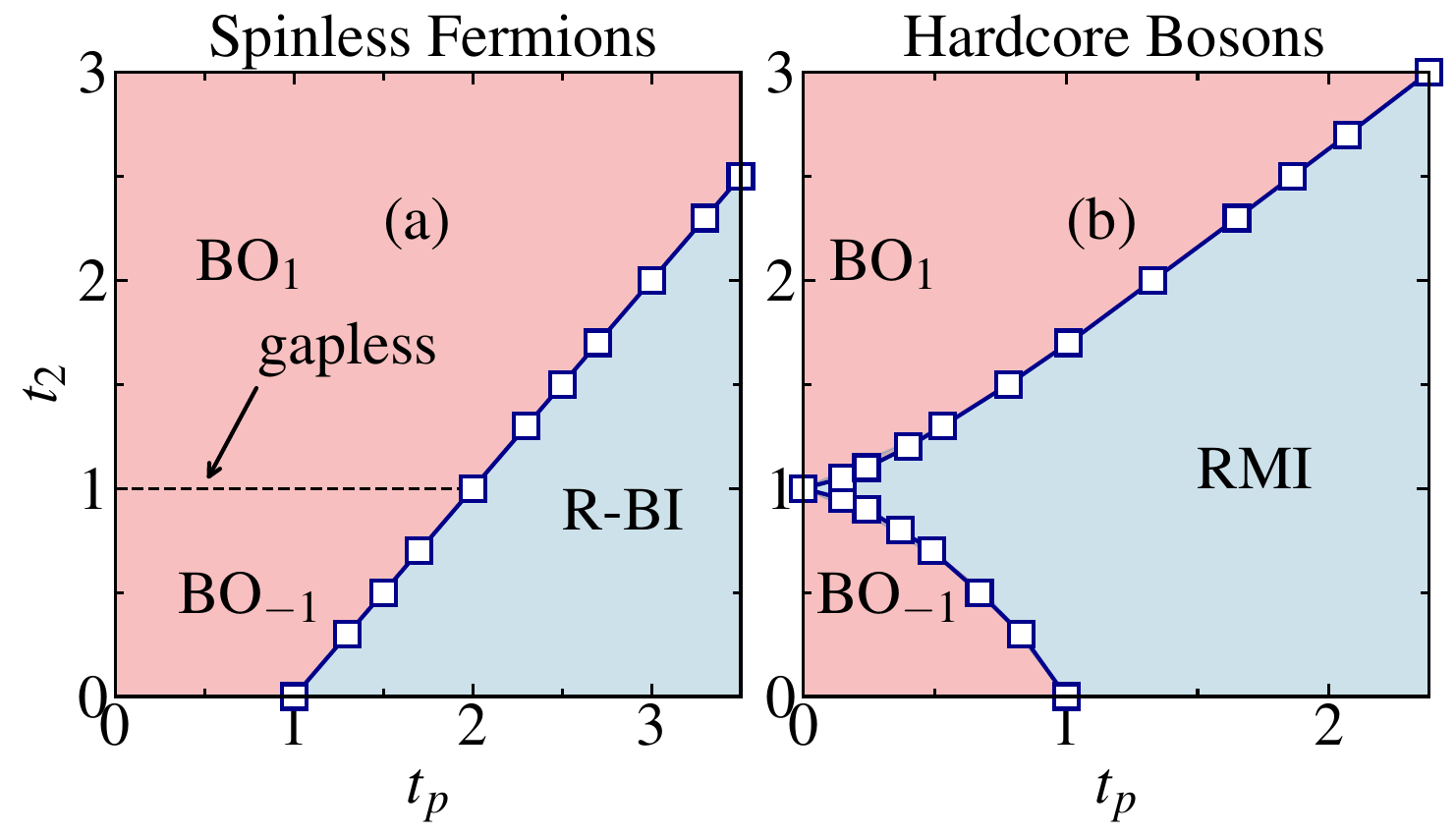}
\end{center}
\caption{The phase diagram in the $t_2-t_p$ plane for spinless fermions and hard-core bosons in (a) and (b) respectively. The red regions denote the topological bond-ordered phases (BO$_1$ and BO$_{-1}$), whereas the blue region indicates the trivial rung insulator phases (R-BI and RMI). The phase boundaries marked by the blue squares demarcate the topological-to-trivial phase transitions in both the panels. The black dashed line in (a) is the gapless boundary between the two distinct topological phases which is absent in (b).}
\label{fig:phase_diagram}
\end{figure}

\paragraph*{Quantum phases.-}
We first discuss the limiting cases of the model considered in order to gain some insights into the topological phases involved.
In the limiting cases $t_1=0$ or $t_2=0$, the ladder reduces to an effective one-dimensional dimerized chain, rendering spinless fermions and hard-core bosons exactly equivalent.
However, when both $t_1$ and $t_2$ are finite, the system acquires a ladder geometry, and the particle statistics becomes crucial.
Previous in-depth studies of such a ladder system have shown that the topological phase survives finite rung hopping before undergoing a transition to a trivial phase beyond a critical value of $t_p$~\cite{nersesyan_ssh_ladder,smitha_ssh_ladder,padhan_ladder}. 
Here we find that a well-defined topological phase transition happens as a function of rung coupling by connecting one topological SSH chain to a trivial SSH chain. The resulting ground-state phase diagrams at half-filling, are shown in Figs.~\ref{fig:phase_diagram}(a) and (b) for spinless fermions and hard-core bosons, respectively.
The phase diagram reveals three gapped phases, among which two are bond-order (BO) phases (showing finite bond kinetic energy oscillations along the legs)~\cite{padhan_ladder,rajashri_ladder} that are adiabatically connected to the phase shown in Figs.~\ref{fig:cartoon}(b) and (c) and are topological in nature. 
The third gapped phase is a rung insulator which has dominant bond kinetic energy along the rungs~\cite{rmi_motrunich,rmi_fabrizio,rmi_laflorencie}, and is adiabatically connected to the phase shown in Fig.~\ref{fig:cartoon}(c). 
While the rung insulator is a band insulator (R-BI) for fermions~\cite{rmi_motrunich,rmi_fabrizio,rmi_laflorencie,rbi_parida_coupling}, it is a Mott insulator (RMI)~\cite{rmi_motrunich,rmi_fabrizio,rmi_laflorencie} for the case of hard-core bosons. Though for both insulators, one particle is dimerized in a rung, the fundamental mechanism governing this property for both fermions and bosons is distinct. The R-BI phase for fermions is a consequence of band filling, where the occupied single-particle states are separated from the unoccupied band by a band gap. However, for the case of bosons, the gap originates solely because of interactions, which suppress density fluctuations at commensurate fillings.
\paragraph{Particle Statistics.-} The striking difference between the two cases lies in the phase separating the two topological BO phases.
While spinless fermions exhibit a gapless line separating the two topological phases [Fig.~\ref{fig:phase_diagram}(a)], hard-core bosons instead develop an intervening RMI phase [Fig.~\ref{fig:phase_diagram}(b)]. At $t_2=0$, both systems undergo the same topological transition at $t_p=1$, consistent with their equivalence in the one-dimensional limit. 
However, upon introducing a finite $t_2$, the critical value of $t_p$ becomes distinct for the two cases.
For the spinless fermionic case, the strength of $t_p$ at which the topological phase transition occurs always increases with the increase in the value of $t_2$, further widening the topological region for any finite value of dimerization strength.
In contrast, for hard-core bosons the topological region that exists at $t_2=0$ (one-dimensional chain limit) starts shrinking as we increase the strength of $t_2$, and disappears 
in the absence of dimerization.
The topological phase then reemerges with a further increase in $t_2$, i.e., with further introduction of dimerization (See Fig.~\ref{fig:phase_diagram}(b)). 
In the absence of dimerization, for spinless fermions the gapless phase remains stable up to a finite $t_p=2$ (black dashed line in Fig.~\ref{fig:phase_diagram}(a)), beyond which a transition to a trivial insulator occurs. On the other hand, any finite $t_p$ immediately drives hard-core bosons from a gapless superfluid phase into a trivial RMI phase [Fig.~\ref{fig:phase_diagram}(b)].
The RMI-to-gapless transition at $t_p=0$ is of the Berezinskii-Kosterlitz-Thouless type, with the charge gap opening exponentially slowly for finite $t_p$~\cite{rmi_fabrizio,rmi_laflorencie}.
In the following, we provide an understanding of these phase transition lines (given by blue squares in Fig.~\ref{fig:phase_diagram}), which gives rise to this crucial difference between the two cases.

\paragraph{Understanding the gap-closing scenarios.-} To understand the origin of these distinct phase boundaries, we first consider the spinless fermionic case, whose phase diagram can be obtained entirely from the single-particle spectrum. The Hamiltonian yields the dispersion relation
\begin{align}
    &E^2=t_p^2 + \frac{1}{2}(|\mu_1|^2+|\mu_2|^2)\nonumber\\&\pm\frac{1}{2}\sqrt{(2t_p^2+|\mu_1|^2+|\mu_2|^2)^2+4(\mu_1\mu_2-t_p^2)(t_p^2-\mu_1^*\mu_2^*)},\nonumber\\
    \label{eq:dispersion}
\end{align}
where $\mu_1=-t_1-t_2e^{-ik}$ and $\mu_2=-t_2-t_1e^{ik}$, with $k$ denoting the crystal momentum. The gap-closing conditions obtained from Eq.~(\ref{eq:dispersion}) reproduce all the phase boundaries in Fig.~\ref{fig:phase_diagram}(a) for spinless fermions and are found to be at $t_p=t_1+t_2$ for $k=0$. 
For the case of hard-core bosons, the understanding of these lines goes as follows.
There are two gap-closing lines: one going from $(t_p,t_2)=(1,0)$ to $(0,1)$ and the other going from
$(0,1)$ to $(\infty,\infty)$ along some direction (see Fig.~\ref{fig:phase_diagram}(b)).
However, it is sufficient
to study the lower line going from $(1,0)$ to $(0,1)$ as the two lines are related by the following symmetry.
Under an exchange of the upper and lower legs, we have $t_1 \leftrightarrow 
t_2$. 
Setting $(t_1,t_2,t_p) = (1,t_2,t_p)$, we can rescale it to $(1/t_2,1,t_p/t_2)$ (since an overall scaling of the Hamiltonian by $1/t_2$ does not change the phase of the system), and then use the exchange symmetry to transform this to $(1,1/t_2,t_p/t_2)$. 
Hence, the system must be in the same phase at the points $(1,t_2,t_p)$ and $(1,1/t_2,t_p/t_2)$.

Next, using the Hartree-Fock approximation in the limit of $t_p=1$ and $t_2\xrightarrow{}0$, we find that the effective hopping amplitude changes from $t_2$ to $-\frac{4}{\pi^2} t_2$ for the bosonic case (see End Matter).
The negative slope indicates that the bending of the gap closing line in the limit of $t_p=1$ and $t_2\xrightarrow{}0$ is opposite to that of the spinless fermionic case.
For spinless fermions, the gap-closing line is given by $t_1+t_2$ with slope towards the right (see Fig.~\ref{fig:phase_diagram}(a)), whereas for the hard-core bosons, the gap-closing line is given by $t_1-\frac{4}{\pi^2}t_2$, in the limit of $t_2\xrightarrow{}0$, which indicates a bending towards the opposite direction compared to the case of spinless fermions.
Moreover, in the absence of dimerization, any finite $t_p$ opens up a trivial gap for hard-core bosons, which is shown by a bosonization analysis~\cite{rmi_laflorencie}.
The above analysis shows that as the value of $t_2$ is increased, the topological phase that is present at $t_2=0$ starts to shrink. At $t_1=t_2$, the topological phase completely disappears, resulting in a trivial region sandwiched between two topological phases. This gives rise to a sequence of transitions from one topological phase to a trivial phase and thereafter to another topological phase, which we refer to as the phenomenon of reentrant topology in the ground state phase diagram.
In the following, we study the topological phase transitions and the emergence of the reentrant topology for the case of hard-core bosons.

\begin{figure}[t]
\begin{center}
\includegraphics[width=1.\columnwidth]{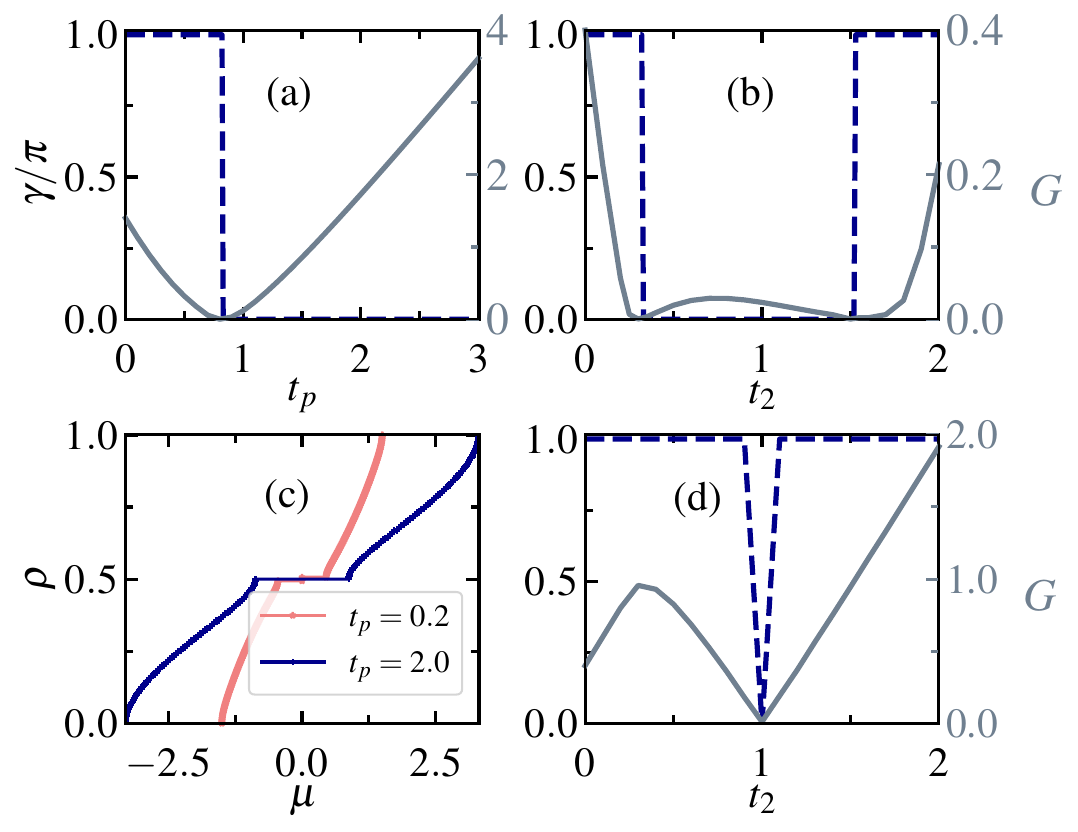}
\end{center}
\caption{Winding number $\gamma/\pi$ (blue dashed) and bulk gap $G$ (gray solid) at half-filling are plotted in (a), (b), and (d). (a) The change in winding number from one to zero indicates the topological-to-trivial phase transition of hard-core bosons driven by $t_p$ for fixed $t_2=0.3$. The gap is finite for all the values of $t_p$ except at the transition point ($t_p\sim 0.83$). (b) The change in winding number from one to zero to one as a function of $t_2$ indicates a reentrant topological phase transition for hard-core bosons for fixed $t_p=0.8$. This behavior is complemented by the gap ($G$). (c) Density $\rho$ as a function of chemical potential $\mu$ for two exemplary values of $t_p$, with fixed $t_2=0.3$, where the plateau indicates the gap in the system and the width of the plateau at $\rho=0.5$ determines the magnitude of the bulk gap at half-filling. (d) Topological-to-topological phase transition of spinless fermions through a gapless critical point.}
\label{fig:invariant}
\end{figure}

\paragraph*{Quantification of the topological phases.-}The topological phases and their transitions as a function of rung coupling $t_p$ can be tracked and characterized by computing the Berry phase $\gamma$ using twisted boundary conditions, where the hopping term at the boundary of a periodic chain is modified as $t \to te^{i\theta}$ for hard-core bosons. 
As $\theta$ varies from $0$ to $2\pi$, the ground state accumulates a phase, given by
\begin{equation}
\gamma ~=~ i~ \int_0^{2\pi} d \theta ~\langle\psi(\theta)|\partial_\theta|\psi(\theta)\rangle.
\label{eq:berry_phase_boson}
\end{equation}
The value of $\gamma$ is $\pi$ if the bulk is topological, whereas it is $0$ if the bulk is trivial. 
Figure~\ref{fig:invariant}(a) shows $\gamma/\pi$  as a function of $t_p$ computed via ED for $L=10$ rungs along the cut $t_2=0.3$ in the phase diagram of hard-core bosons. 
A sharp jump of $\gamma/\pi$ from $1$ to $0$ near $t_p\sim0.83$ marks the topological phase transition point. 
The topological invariant also agrees well with the presence and absence of edge states in the corresponding regimes (See Fig.~\ref{fig:cartoon} and SM for numerical details).
We find the phase boundaries (blue squares) in Fig.~\ref{fig:phase_diagram}(b) by tracking this transition point through the change of the invariant $\gamma/\pi$ for different values of $t_2$.
For the spinless fermion case, the invariant is calculated from the filled bands as 
\begin{equation}
\gamma = -\mathfrak{I}\ln \prod_{j=0}^{N-1}
\langle u_{k_j}|u_{k_{j+1}}\rangle,
\label{eq:berry_phase_fermion}
\end{equation}
where $\mathfrak{I}$ denotes the imaginary part, $k_j=2\pi j/(Na)$ is the $j$-th momentum point in the Brillouin zone, and $|u_{k_j}\rangle$ is the occupied Bloch eigenstate.
The analytical boundaries obtained from the band structure calculation for free fermions agree well with the topological invariant $\gamma$.
%
%
The above analysis is further complemented by the bulk gap $G$ shown in Fig.~\ref{fig:invariant}.
We plot the extrapolated value of the bulk gap $G$ obtained from the width of the plateau in the density {$\rho$} versus the chemical potential {$\mu$} at half-filling~\cite{rajashri_four_quadrant} as shown in Fig.~\ref{fig:invariant}(c) for the case of hard-core bosons.
For fermions, $G$ is obtained from the energy difference between the $(L/2+1)$-th and $(L/2)$-th eigenstates under PBC.
$G$ is found to vanish at the topological phase transition point and remains finite elsewhere, confirming the gapped topological and trivial phases which agree well with the above topological phase transition.

Intriguingly, at a fixed value of $t_p$, we find two consecutive topological phase transitions as a function of $t_2$ in the case of hard-core bosons, in the form of a reentrant topological phase transition.
Figure~\ref{fig:invariant}(b) shows the topological invariant, bulk gap $G$, and string order $O_{\text{str}}$ along the cut $t_p=0.8$ of Fig.~\ref{fig:phase_diagram}(b). 
The invariant $\gamma$ exhibits two jumps, from $1$ to $0$ and subsequently from $0$ to $1$, signaling a reentrant topological phase transition. 
Consistently, $G$ also closes at both points and remains finite at all other points.
However, such a phenomenon does not occur when one considers spinless fermions instead.
In order to show this, we consider the exemplary cut $t_p=0.8$ in Fig.~\ref{fig:phase_diagram}(a) and calculate $\gamma/\pi$, which is shown in Fig.~\ref{fig:invariant}(d).
The value of $\gamma/\pi$ (blue dashed line) stays quantized at $1$ across the full range of $t_2$, except at $t_1=t_2=1$ where the bulk gap ($G$) closes.
The non-trivial values of $\gamma/\pi$ in the gapped phases on both sides of the gapless point confirm the topological nature of both the phases.
After all the above analysis, we will now characterize this phenomenon by an experimentally relevant quantity called Thouless charge pumping.

\paragraph*{Thouless charge pumping.-}
To further probe the topology, we study Thouless charge pumping (TCP)~\cite{Thouless1983}, which connects topology to quantized transport through a cyclic and adiabatic modulation of the system parameters. 
In the pumping cycle, the system evolves from a topological to a trivial phase and back to the initial topological phase, with the path encircling the transition point while keeping the bulk gap open via an inversion-symmetry-breaking term. 
This process transports an integer number of particles, directly tied to the topological invariant. 
Originally proposed in the context of non-interacting systems, TCP has now been extended to interacting settings and has been proven to be a powerful probe for topological phases, both theoretically and experimentally~\cite{monika_review,bound_pump1,bound_pump3,Kuno2017,bertok_pump,mondal_phonon,Hayward2018,spin_pumping,pumping_quasicrystals,pumping_1d,hubbarad_thouless_pump,qubit_pumping,seba_nphy_pumming,seba_nature_pumping,pumping_esslinger,pump_reversal_esslinger}. 
\begin{figure}[t]
\begin{center}
\includegraphics[width=1.\columnwidth]{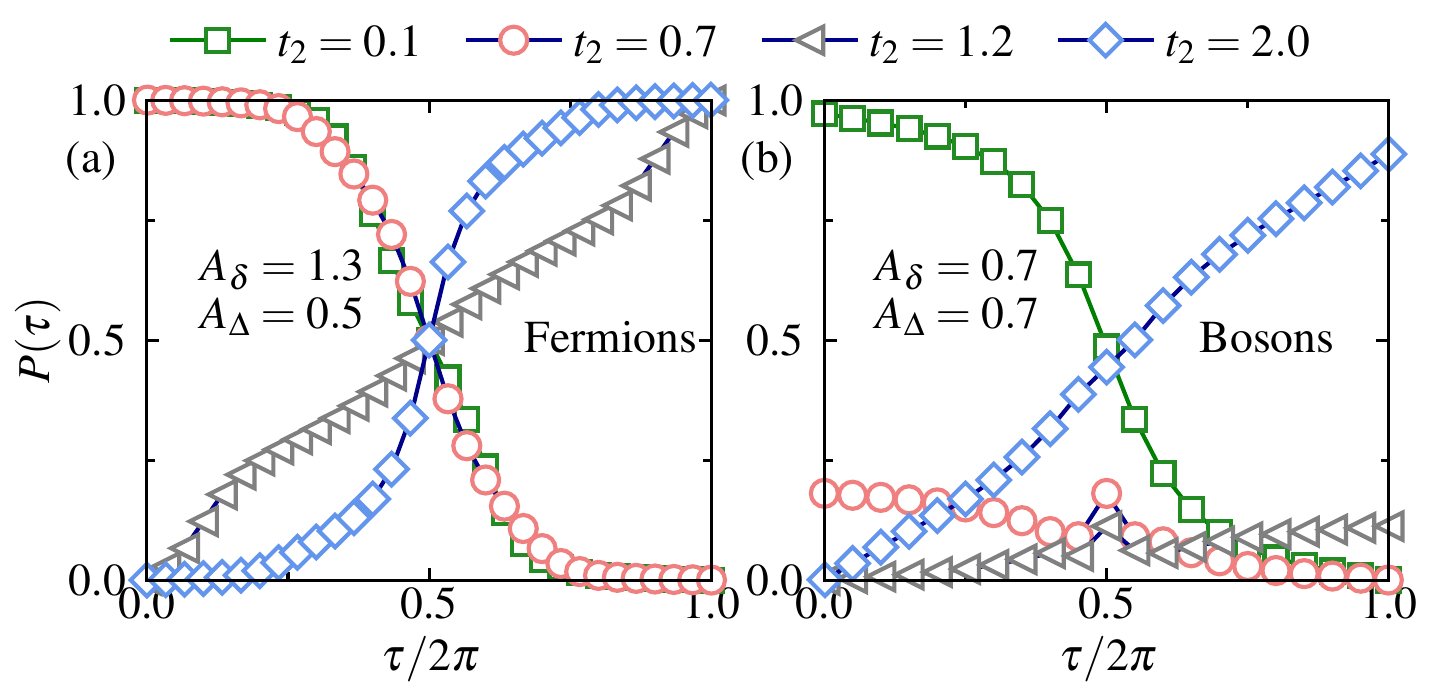}
\end{center}
\caption{Evolution of the polarization $P(\tau)$ as a function of $\tau$ for a system with $L=100$ rungs, shown at four different values of the hopping strengths, i.e., $t_2=0.1$ (green squares), $t_2=0.7$ (red circles), $t_2=1.2$ (gray triangles), and $t_2=2$ (blue diamonds). See text for the details of other system parameters. While spinless fermions show smooth pumping for all values of $t_2$, for hard-core bosons there is a regime of values of $t_2$ where there is no pumping at all. This shows that there is a regime of $t_2$ for which the cycle completely remains in a trivial phase for the case of hard-core bosons.}
\label{fig:pumping}
\end{figure}
In one dimension, the Rice--Mele model~\cite{rice_mele_model,Asboth2016_ssh} defines the pumping protocol which, in our case, is defined by the Hamiltonian
\begin{align}
H_{p}(\tau) = H_{||} ~-~ (t_0 - \delta(\tau)) \sum_i (a_i^\dagger b_i+\text{H.c.}) \nonumber\\
~~~~~~~~ + ~\Delta(\tau)\sum\limits_i [(-1)^{i+1}n_{a,i}+(-1)^{i}n_{b,i}].
\label{eq:pumping_hamiltonian}
\end{align}
Here $\tau$ is the adiabatic pumping parameter, with rung hopping and staggered onsite potential modulated as $\delta(\tau)=A_\delta\cos(2\pi\tau)$ and $\Delta(\tau)=A_\Delta\sin(2\pi\tau)$.
In Eq.~\eqref{eq:pumping_hamiltonian} $(t_0,\Delta=0)$ defines the origin of the pumping cycle in the $\delta$–$\Delta$ plane, and $n_{\alpha,i}$ with $\alpha = a,b$ represents the corresponding number operator. 
To investigate the TCP, we study four pumping cycles corresponding to different hopping strengths ($t_2$).
We compute the polarization as a function of $\tau$ for a system of $L=100$ rungs, shown in Figs.~\ref{fig:pumping}(a,b).
For hard-core bosons, the polarization is calculated in real space using OBC as
\begin{align}
&P(\tau)=\bigg(\frac{1}{2L}\sum_{i=1}^{L}
\bra{\psi(\tau)}
\left((2i-1)-\frac{2L+1}{2}\right)
n_{a,i}
\ket{\psi(\tau)}\nonumber\\
&~~~~~ +\bra{\psi(\tau)}
\left(2i-\frac{2L+1}{2}\right)
n_{b,i}
\ket{\psi(\tau)}\bigg) ~~(\text{mod}~Q),
\end{align}
where $\ket{\psi(\tau)}$ is the instantaneous ground state, and the lattice constant is set to unity.
The net charge pumped in one cycle is obtained as
$Q=\int_0^1 d\tau\,\partial_\tau P(\tau).$
For spinless fermions, the polarization is computed from the occupied Bloch eigenstate of the Rice--Mele Hamiltonian [Eq.~\eqref{eq:pumping_hamiltonian}] under PBC as~\cite{berry_phase}
\begin{equation}
P=-\frac{1}{2\pi}\mathfrak{I}\ln
\prod_{j=0}^{N-1}
\langle u_{k_j}|u_{k_{j+1}}\rangle
\quad(\mathrm{mod}\;Q),
\label{eq:polarization}
\end{equation}
where $k_j=2\pi j/N$ labels the discrete momentum points in the Brillouin zone, and $|u_{k_j}\rangle$ is the occupied Bloch eigenstate of
$H(k)=
\begin{pmatrix}
\Delta(\tau) & h(k,\tau)\\
h^{*}(k,\tau) & -\Delta(\tau)
\end{pmatrix}$,
where
$h(k,\tau)=-t_p(\tau)-t_2e^{-ik}-t_1e^{ik},$
with $t_p(\tau)=t_0+\delta(\tau)$.
For all values of $t_2$, the polarization evolves smoothly during the pumping cycle, yielding a quantized pumped charge $|Q|=1$ for spinless fermions [Fig.~\ref{fig:pumping}(a)], since the pumping loop encloses the topological transition point for any finite hopping dimerization.
However, the pumping direction reverses for $t_1<t_2$ compared to $t_1>t_2$. For example, at $t_2=0.1$, the polarization evolves smoothly from $1$ to $0$, whereas for $t_2=2$, it evolves from $0$ to $1$. Based on this reversal, we denote the two topological phases as BO$_{-1}$ ($t_1<t_2$) and BO$_1$ ($t_1>t_2$) respectively.
While hard-core bosons also exhibit analogous quantized pumping for $t_2=0.1$ and $2$, a sharp contrast appears at $t_2=0.7$ and $1.2$ [Fig.~\ref{fig:phase_diagram}(b)], where the polarization becomes discontinuous, indicating that the pumping loop lies entirely in the trivial phase. This confirms the existence of a finite trivial region separating the two topological phases, thereby establishing the reentrant topology of the hard-core boson system. 
%


\paragraph*{Outlook.-}
\label{sec:conc}
We have shown that coupling two SSH chains with opposite dimerization patterns in a two-leg ladder geometry gives rise to strikingly different topological behaviors for spinless fermions and hard-core bosons. 
%
For hard-core bosons, we uncovered a qualitatively distinct phase diagram where a trivial phase is sandwiched between two topological phases with different characters, resulting in a reentrant topological phase transition.
In contrast, for fermions, the two topological phases are separated by a gapless line rather than an intervening trivial phase, highlighting the crucial role of particle statistics away from strictly one dimension.
Such ladder geometries with opposite dimerization patterns have already been realized experimentally~\cite{expt_ssh_ladder}, providing a concrete platform for investigating the distinct topological phenomena revealed here.
Moreover, a natural next step would be to identify systems where changing the underlying particle statistics can qualitatively alter, or even completely destroy, the topological phases present in the corresponding fermionic or bosonic counterparts.
Having gleaned the dramatic effect of quantum statistics on topological phases even in the simple ladder system, extensions full-fledged higher-dimensional systems would open up many rich prospects, such as its effects on intrinsically ordered topological phases, its role in these settings on emergent fractionalization, and its interplay with frustration and long-range interactions.

\paragraph{Acknowledgments.-} 
T.M. acknowledges support from Anusandhan National Research Foundation (ANRF), Government of India, through project No. ANRF/ARG/2025/007442/PS.

\bibliography{ref}

\begin{thebibliography}{49}%
\makeatletter
\providecommand \@ifxundefined [1]{%
 \@ifx{#1\undefined}
}%
\providecommand \@ifnum [1]{%
 \ifnum #1\expandafter \@firstoftwo
 \else \expandafter \@secondoftwo
 \fi
}%
\providecommand \@ifx [1]{%
 \ifx #1\expandafter \@firstoftwo
 \else \expandafter \@secondoftwo
 \fi
}%
\providecommand \natexlab [1]{#1}%
\providecommand \enquote  [1]{``#1''}%
\providecommand \bibnamefont  [1]{#1}%
\providecommand \bibfnamefont [1]{#1}%
\providecommand \citenamefont [1]{#1}%
\providecommand \href@noop [0]{\@secondoftwo}%
\providecommand \href [0]{\begingroup \@sanitize@url \@href}%
\providecommand \@href[1]{\@@startlink{#1}\@@href}%
\providecommand \@@href[1]{\endgroup#1\@@endlink}%
\providecommand \@sanitize@url [0]{\catcode `\\12\catcode `\$12\catcode `\&12\catcode `\#12\catcode `\^12\catcode `\_12\catcode `\%12\relax}%
\providecommand \@@startlink[1]{}%
\providecommand \@@endlink[0]{}%
\providecommand \url  [0]{\begingroup\@sanitize@url \@url }%
\providecommand \@url [1]{\endgroup\@href {#1}{\urlprefix }}%
\providecommand \urlprefix  [0]{URL }%
\providecommand \Eprint [0]{\href }%
\providecommand \doibase [0]{https://doi.org/}%
\providecommand \selectlanguage [0]{\@gobble}%
\providecommand \bibinfo  [0]{\@secondoftwo}%
\providecommand \bibfield  [0]{\@secondoftwo}%
\providecommand \translation [1]{[#1]}%
\providecommand \BibitemOpen [0]{}%
\providecommand \bibitemStop [0]{}%
\providecommand \bibitemNoStop [0]{.\EOS\space}%
\providecommand \EOS [0]{\spacefactor3000\relax}%
\providecommand \BibitemShut  [1]{\csname bibitem#1\endcsname}%
\let\auto@bib@innerbib\@empty
\bibitem [{\citenamefont {Rigol}\ and\ \citenamefont {Muramatsu}(2005)}]{Rigol2005}%
  \BibitemOpen
  \bibfield  {author} {\bibinfo {author} {\bibfnamefont {M.}~\bibnamefont {Rigol}}\ and\ \bibinfo {author} {\bibfnamefont {A.}~\bibnamefont {Muramatsu}},\ }\bibfield  {title} {\bibinfo {title} {Hard-core bosons and spinless fermions trapped on 1d lattices},\ }\href {https://doi.org/10.1007/s10909-005-2274-3} {\bibfield  {journal} {\bibinfo  {journal} {Journal of Low Temperature Physics}\ }\textbf {\bibinfo {volume} {138}},\ \bibinfo {pages} {645} (\bibinfo {year} {2005})}\BibitemShut {NoStop}%
\bibitem [{\citenamefont {Cheong}\ and\ \citenamefont {Henley}(2009)}]{cheong_henley}%
  \BibitemOpen
  \bibfield  {author} {\bibinfo {author} {\bibfnamefont {S.-A.}\ \bibnamefont {Cheong}}\ and\ \bibinfo {author} {\bibfnamefont {C.~L.}\ \bibnamefont {Henley}},\ }\bibfield  {title} {\bibinfo {title} {Exact ground states and correlation functions of chain and ladder models of interacting hardcore bosons or spinless fermions},\ }\href {https://doi.org/10.1103/PhysRevB.80.165124} {\bibfield  {journal} {\bibinfo  {journal} {Phys. Rev. B}\ }\textbf {\bibinfo {volume} {80}},\ \bibinfo {pages} {165124} (\bibinfo {year} {2009})}\BibitemShut {NoStop}%
\bibitem [{\citenamefont {Girardeau}(1960)}]{Girardeau_jmp}%
  \BibitemOpen
  \bibfield  {author} {\bibinfo {author} {\bibfnamefont {M.}~\bibnamefont {Girardeau}},\ }\bibfield  {title} {\bibinfo {title} {Relationship between systems of impenetrable bosons and fermions in one dimension},\ }\href {https://doi.org/10.1063/1.1703687} {\bibfield  {journal} {\bibinfo  {journal} {Journal of Mathematical Physics}\ }\textbf {\bibinfo {volume} {1}},\ \bibinfo {pages} {516} (\bibinfo {year} {1960})}\BibitemShut {NoStop}%
\bibitem [{\citenamefont {Jordan}\ and\ \citenamefont {Wigner}(1928)}]{Jordan1928}%
  \BibitemOpen
  \bibfield  {author} {\bibinfo {author} {\bibfnamefont {P.}~\bibnamefont {Jordan}}\ and\ \bibinfo {author} {\bibfnamefont {E.}~\bibnamefont {Wigner}},\ }\bibfield  {title} {\bibinfo {title} {{\"U}ber das paulische {\"a}quivalenzverbot},\ }\href {https://doi.org/10.1007/BF01331938} {\bibfield  {journal} {\bibinfo  {journal} {Zeitschrift f{\"u}r Physik}\ }\textbf {\bibinfo {volume} {47}},\ \bibinfo {pages} {631} (\bibinfo {year} {1928})}\BibitemShut {NoStop}%
\bibitem [{\citenamefont {Su}\ \emph {et~al.}(1979)\citenamefont {Su}, \citenamefont {Schrieffer},\ and\ \citenamefont {Heeger}}]{ssh_model}%
  \BibitemOpen
  \bibfield  {author} {\bibinfo {author} {\bibfnamefont {W.~P.}\ \bibnamefont {Su}}, \bibinfo {author} {\bibfnamefont {J.~R.}\ \bibnamefont {Schrieffer}},\ and\ \bibinfo {author} {\bibfnamefont {A.~J.}\ \bibnamefont {Heeger}},\ }\bibfield  {title} {\bibinfo {title} {Solitons in polyacetylene},\ }\href {https://doi.org/10.1103/PhysRevLett.42.1698} {\bibfield  {journal} {\bibinfo  {journal} {Phys. Rev. Lett.}\ }\textbf {\bibinfo {volume} {42}},\ \bibinfo {pages} {1698} (\bibinfo {year} {1979})}\BibitemShut {NoStop}%
\bibitem [{\citenamefont {Asb{\'o}th}\ \emph {et~al.}(2016)\citenamefont {Asb{\'o}th}, \citenamefont {Oroszl{\'a}ny},\ and\ \citenamefont {P{\'a}lyi}}]{Asboth2016_ssh}%
  \BibitemOpen
  \bibfield  {author} {\bibinfo {author} {\bibfnamefont {J.~K.}\ \bibnamefont {Asb{\'o}th}}, \bibinfo {author} {\bibfnamefont {L.}~\bibnamefont {Oroszl{\'a}ny}},\ and\ \bibinfo {author} {\bibfnamefont {A.}~\bibnamefont {P{\'a}lyi}},\ }\bibinfo {title} {The {S}u-{S}chrieffer-{H}eeger model},\ in\ \href {https://doi.org/10.1007/978-3-319-25607-8_1} {\emph {\bibinfo {booktitle} {A Short Course on Topological Insulators}}}\ (\bibinfo  {publisher} {Springer International Publishing},\ \bibinfo {address} {Cham},\ \bibinfo {year} {2016})\ pp.\ \bibinfo {pages} {1--22}\BibitemShut {NoStop}%
\bibitem [{\citenamefont {Atala}\ \emph {et~al.}(2013)\citenamefont {Atala}, \citenamefont {Aidelsburger}, \citenamefont {Barreiro}, \citenamefont {Abanin}, \citenamefont {Kitagawa}, \citenamefont {Demler},\ and\ \citenamefont {Bloch}}]{Atala2013}%
  \BibitemOpen
  \bibfield  {author} {\bibinfo {author} {\bibfnamefont {M.}~\bibnamefont {Atala}}, \bibinfo {author} {\bibfnamefont {M.}~\bibnamefont {Aidelsburger}}, \bibinfo {author} {\bibfnamefont {J.~T.}\ \bibnamefont {Barreiro}}, \bibinfo {author} {\bibfnamefont {D.}~\bibnamefont {Abanin}}, \bibinfo {author} {\bibfnamefont {T.}~\bibnamefont {Kitagawa}}, \bibinfo {author} {\bibfnamefont {E.}~\bibnamefont {Demler}},\ and\ \bibinfo {author} {\bibfnamefont {I.}~\bibnamefont {Bloch}},\ }\bibfield  {title} {\bibinfo {title} {Direct measurement of the zak phase in topological bloch bands},\ }\href {https://doi.org/https://doi.org/10.1038/nphys2790} {\bibfield  {journal} {\bibinfo  {journal} {Nature Physics}\ }\textbf {\bibinfo {volume} {9}},\ \bibinfo {pages} {795} (\bibinfo {year} {2013})}\BibitemShut {NoStop}%
\bibitem [{\citenamefont {Nakajima}\ \emph {et~al.}(2016)\citenamefont {Nakajima}, \citenamefont {Tomita}, \citenamefont {Taie}, \citenamefont {Ichinose}, \citenamefont {Ozawa}, \citenamefont {Wang}, \citenamefont {Troyer},\ and\ \citenamefont {Takahashi}}]{Takahashi2016pumping}%
  \BibitemOpen
  \bibfield  {author} {\bibinfo {author} {\bibfnamefont {S.}~\bibnamefont {Nakajima}}, \bibinfo {author} {\bibfnamefont {T.}~\bibnamefont {Tomita}}, \bibinfo {author} {\bibfnamefont {S.}~\bibnamefont {Taie}}, \bibinfo {author} {\bibfnamefont {T.}~\bibnamefont {Ichinose}}, \bibinfo {author} {\bibfnamefont {H.}~\bibnamefont {Ozawa}}, \bibinfo {author} {\bibfnamefont {L.}~\bibnamefont {Wang}}, \bibinfo {author} {\bibfnamefont {M.}~\bibnamefont {Troyer}},\ and\ \bibinfo {author} {\bibfnamefont {Y.}~\bibnamefont {Takahashi}},\ }\bibfield  {title} {\bibinfo {title} {Topological {T}houless pumping of ultracold fermions},\ }\href {https://doi.org/http://dx.doi.org/10.1038/nphys3622 10.1038/nphys3622} {\bibfield  {journal} {\bibinfo  {journal} {Nature Physics}\ }\textbf {\bibinfo {volume} {12}},\ \bibinfo {pages} {296} (\bibinfo {year} {2016})}\BibitemShut {NoStop}%
\bibitem [{\citenamefont {Lohse}\ \emph {et~al.}(2016)\citenamefont {Lohse}, \citenamefont {Schweizer}, \citenamefont {Zilberberg}, \citenamefont {Aidelsburger},\ and\ \citenamefont {Bloch}}]{Lohse2016}%
  \BibitemOpen
  \bibfield  {author} {\bibinfo {author} {\bibfnamefont {M.}~\bibnamefont {Lohse}}, \bibinfo {author} {\bibfnamefont {C.}~\bibnamefont {Schweizer}}, \bibinfo {author} {\bibfnamefont {O.}~\bibnamefont {Zilberberg}}, \bibinfo {author} {\bibfnamefont {M.}~\bibnamefont {Aidelsburger}},\ and\ \bibinfo {author} {\bibfnamefont {I.}~\bibnamefont {Bloch}},\ }\bibfield  {title} {\bibinfo {title} {A {T}houless quantum pump with ultracold bosonic atoms in an optical superlattice},\ }\href {https://doi.org/10.1038/nphys3584} {\bibfield  {journal} {\bibinfo  {journal} {Nature Physics}\ }\textbf {\bibinfo {volume} {12}},\ \bibinfo {pages} {350} (\bibinfo {year} {2016})}\BibitemShut {NoStop}%
\bibitem [{\citenamefont {Mukherjee}\ \emph {et~al.}(2017)\citenamefont {Mukherjee}, \citenamefont {Spracklen}, \citenamefont {Valiente}, \citenamefont {Andersson}, \citenamefont {{\"O}hberg}, \citenamefont {Goldman},\ and\ \citenamefont {Thomson}}]{Mukherjee2017}%
  \BibitemOpen
  \bibfield  {author} {\bibinfo {author} {\bibfnamefont {S.}~\bibnamefont {Mukherjee}}, \bibinfo {author} {\bibfnamefont {A.}~\bibnamefont {Spracklen}}, \bibinfo {author} {\bibfnamefont {M.}~\bibnamefont {Valiente}}, \bibinfo {author} {\bibfnamefont {E.}~\bibnamefont {Andersson}}, \bibinfo {author} {\bibfnamefont {P.}~\bibnamefont {{\"O}hberg}}, \bibinfo {author} {\bibfnamefont {N.}~\bibnamefont {Goldman}},\ and\ \bibinfo {author} {\bibfnamefont {R.~R.}\ \bibnamefont {Thomson}},\ }\bibfield  {title} {\bibinfo {title} {Experimental observation of anomalous topological edge modes in a slowly driven photonic lattice},\ }\href {https://doi.org/10.1038/ncomms13918} {\bibfield  {journal} {\bibinfo  {journal} {Nature Communications}\ }\textbf {\bibinfo {volume} {8}},\ \bibinfo {pages} {13918} (\bibinfo {year} {2017})}\BibitemShut {NoStop}%
\bibitem [{\citenamefont {Lu}\ \emph {et~al.}(2014)\citenamefont {Lu}, \citenamefont {Joannopoulos},\ and\ \citenamefont {Solja{\v{c}}i{\'{c}}}}]{Lu2014}%
  \BibitemOpen
  \bibfield  {author} {\bibinfo {author} {\bibfnamefont {L.}~\bibnamefont {Lu}}, \bibinfo {author} {\bibfnamefont {J.~D.}\ \bibnamefont {Joannopoulos}},\ and\ \bibinfo {author} {\bibfnamefont {M.}~\bibnamefont {Solja{\v{c}}i{\'{c}}}},\ }\bibfield  {title} {\bibinfo {title} {Topological photonics},\ }\href {https://doi.org/10.1038/nphoton.2014.248} {\bibfield  {journal} {\bibinfo  {journal} {Nature Photonics}\ }\textbf {\bibinfo {volume} {8}},\ \bibinfo {pages} {821} (\bibinfo {year} {2014})}\BibitemShut {NoStop}%
\bibitem [{\citenamefont {Thatcher}\ \emph {et~al.}(2022)\citenamefont {Thatcher}, \citenamefont {Fairfield}, \citenamefont {Merlo-Ramírez},\ and\ \citenamefont {Merlo}}]{ssh_expt_1}%
  \BibitemOpen
  \bibfield  {author} {\bibinfo {author} {\bibfnamefont {L.}~\bibnamefont {Thatcher}}, \bibinfo {author} {\bibfnamefont {P.}~\bibnamefont {Fairfield}}, \bibinfo {author} {\bibfnamefont {L.}~\bibnamefont {Merlo-Ramírez}},\ and\ \bibinfo {author} {\bibfnamefont {J.~M.}\ \bibnamefont {Merlo}},\ }\bibfield  {title} {\bibinfo {title} {Experimental observation of topological phase transitions in a mechanical 1d-ssh model},\ }\href {https://doi.org/10.1088/1402-4896/ac4ed2} {\bibfield  {journal} {\bibinfo  {journal} {Physica Scripta}\ }\textbf {\bibinfo {volume} {97}},\ \bibinfo {pages} {035702} (\bibinfo {year} {2022})}\BibitemShut {NoStop}%
\bibitem [{\citenamefont {Meier}\ \emph {et~al.}(2016)\citenamefont {Meier}, \citenamefont {An},\ and\ \citenamefont {Gadway}}]{ssh_expt_2}%
  \BibitemOpen
  \bibfield  {author} {\bibinfo {author} {\bibfnamefont {E.~J.}\ \bibnamefont {Meier}}, \bibinfo {author} {\bibfnamefont {F.~A.}\ \bibnamefont {An}},\ and\ \bibinfo {author} {\bibfnamefont {B.}~\bibnamefont {Gadway}},\ }\bibfield  {title} {\bibinfo {title} {Observation of the topological soliton state in the {S}u--{S}chrieffer--{H}eeger model},\ }\href {https://doi.org/10.1038/ncomms13986} {\bibfield  {journal} {\bibinfo  {journal} {Nature Communications}\ }\textbf {\bibinfo {volume} {7}},\ \bibinfo {pages} {13986} (\bibinfo {year} {2016})}\BibitemShut {NoStop}%
\bibitem [{\citenamefont {de~Léséleuc}\ \emph {et~al.}(2019)\citenamefont {de~Léséleuc}, \citenamefont {Lienhard}, \citenamefont {Scholl}, \citenamefont {Barredo}, \citenamefont {Weber}, \citenamefont {Lang}, \citenamefont {Büchler}, \citenamefont {Lahaye},\ and\ \citenamefont {Browaeys}}]{browyes}%
  \BibitemOpen
  \bibfield  {author} {\bibinfo {author} {\bibfnamefont {S.}~\bibnamefont {de~Léséleuc}}, \bibinfo {author} {\bibfnamefont {V.}~\bibnamefont {Lienhard}}, \bibinfo {author} {\bibfnamefont {P.}~\bibnamefont {Scholl}}, \bibinfo {author} {\bibfnamefont {D.}~\bibnamefont {Barredo}}, \bibinfo {author} {\bibfnamefont {S.}~\bibnamefont {Weber}}, \bibinfo {author} {\bibfnamefont {N.}~\bibnamefont {Lang}}, \bibinfo {author} {\bibfnamefont {H.~P.}\ \bibnamefont {Büchler}}, \bibinfo {author} {\bibfnamefont {T.}~\bibnamefont {Lahaye}},\ and\ \bibinfo {author} {\bibfnamefont {A.}~\bibnamefont {Browaeys}},\ }\bibfield  {title} {\bibinfo {title} {Observation of a symmetry-protected topological phase of interacting bosons with rydberg atoms},\ }\href {https://doi.org/10.1126/science.aav9105} {\bibfield  {journal} {\bibinfo  {journal} {Science}\ }\textbf {\bibinfo {volume} {365}},\ \bibinfo {pages} {775} (\bibinfo {year} {2019})}\BibitemShut {NoStop}%
\bibitem [{\citenamefont {Liu}\ \emph {et~al.}(2022)\citenamefont {Liu}, \citenamefont {Cao}, \citenamefont {Chen}, \citenamefont {Wang}, \citenamefont {Yang},\ and\ \citenamefont {Zhang}}]{EC1}%
  \BibitemOpen
  \bibfield  {author} {\bibinfo {author} {\bibfnamefont {Y.}~\bibnamefont {Liu}}, \bibinfo {author} {\bibfnamefont {W.}~\bibnamefont {Cao}}, \bibinfo {author} {\bibfnamefont {W.}~\bibnamefont {Chen}}, \bibinfo {author} {\bibfnamefont {H.}~\bibnamefont {Wang}}, \bibinfo {author} {\bibfnamefont {L.}~\bibnamefont {Yang}},\ and\ \bibinfo {author} {\bibfnamefont {X.}~\bibnamefont {Zhang}},\ }\bibfield  {title} {\bibinfo {title} {Fully integrated topological electronics},\ }\href {https://doi.org/10.1038/s41598-022-17010-8} {\bibfield  {journal} {\bibinfo  {journal} {Scientific Reports}\ }\textbf {\bibinfo {volume} {12}},\ \bibinfo {pages} {13410} (\bibinfo {year} {2022})}\BibitemShut {NoStop}%
\bibitem [{\citenamefont {Lee}\ \emph {et~al.}(2018)\citenamefont {Lee}, \citenamefont {Imhof}, \citenamefont {Berger}, \citenamefont {Bayer}, \citenamefont {Brehm}, \citenamefont {Molenkamp}, \citenamefont {Kiessling},\ and\ \citenamefont {Thomale}}]{EC2}%
  \BibitemOpen
  \bibfield  {author} {\bibinfo {author} {\bibfnamefont {C.~H.}\ \bibnamefont {Lee}}, \bibinfo {author} {\bibfnamefont {S.}~\bibnamefont {Imhof}}, \bibinfo {author} {\bibfnamefont {C.}~\bibnamefont {Berger}}, \bibinfo {author} {\bibfnamefont {F.}~\bibnamefont {Bayer}}, \bibinfo {author} {\bibfnamefont {J.}~\bibnamefont {Brehm}}, \bibinfo {author} {\bibfnamefont {L.~W.}\ \bibnamefont {Molenkamp}}, \bibinfo {author} {\bibfnamefont {T.}~\bibnamefont {Kiessling}},\ and\ \bibinfo {author} {\bibfnamefont {R.}~\bibnamefont {Thomale}},\ }\bibfield  {title} {\bibinfo {title} {Topolectrical circuits},\ }\href {https://doi.org/10.1038/s42005-018-0035-2} {\bibfield  {journal} {\bibinfo  {journal} {Communications Physics}\ }\textbf {\bibinfo {volume} {1}},\ \bibinfo {pages} {39} (\bibinfo {year} {2018})}\BibitemShut {NoStop}%
\bibitem [{\citenamefont {White}(1992)}]{white1992}%
  \BibitemOpen
  \bibfield  {author} {\bibinfo {author} {\bibfnamefont {S.~R.}\ \bibnamefont {White}},\ }\bibfield  {title} {\bibinfo {title} {Density matrix formulation for quantum renormalization groups},\ }\href {https://doi.org/10.1103/PhysRevLett.69.2863} {\bibfield  {journal} {\bibinfo  {journal} {Phys. Rev. Lett.}\ }\textbf {\bibinfo {volume} {69}},\ \bibinfo {pages} {2863} (\bibinfo {year} {1992})}\BibitemShut {NoStop}%
\bibitem [{\citenamefont {Schollw\"ock}(2005)}]{schollowck_dmrg_rev}%
  \BibitemOpen
  \bibfield  {author} {\bibinfo {author} {\bibfnamefont {U.}~\bibnamefont {Schollw\"ock}},\ }\bibfield  {title} {\bibinfo {title} {The density-matrix renormalization group},\ }\href {https://doi.org/10.1103/RevModPhys.77.259} {\bibfield  {journal} {\bibinfo  {journal} {Rev. Mod. Phys.}\ }\textbf {\bibinfo {volume} {77}},\ \bibinfo {pages} {259} (\bibinfo {year} {2005})}\BibitemShut {NoStop}%
\bibitem [{\citenamefont {Cirac}\ \emph {et~al.}(2021)\citenamefont {Cirac}, \citenamefont {P\'erez-Garc\'{\i}a}, \citenamefont {Schuch},\ and\ \citenamefont {Verstraete}}]{Verstraete_rev}%
  \BibitemOpen
  \bibfield  {author} {\bibinfo {author} {\bibfnamefont {J.~I.}\ \bibnamefont {Cirac}}, \bibinfo {author} {\bibfnamefont {D.}~\bibnamefont {P\'erez-Garc\'{\i}a}}, \bibinfo {author} {\bibfnamefont {N.}~\bibnamefont {Schuch}},\ and\ \bibinfo {author} {\bibfnamefont {F.}~\bibnamefont {Verstraete}},\ }\bibfield  {title} {\bibinfo {title} {Matrix product states and projected entangled pair states: Concepts, symmetries, theorems},\ }\href {https://doi.org/10.1103/RevModPhys.93.045003} {\bibfield  {journal} {\bibinfo  {journal} {Rev. Mod. Phys.}\ }\textbf {\bibinfo {volume} {93}},\ \bibinfo {pages} {045003} (\bibinfo {year} {2021})}\BibitemShut {NoStop}%
\bibitem [{\citenamefont {Schollwöck}(2011)}]{schollowck_mps}%
  \BibitemOpen
  \bibfield  {author} {\bibinfo {author} {\bibfnamefont {U.}~\bibnamefont {Schollwöck}},\ }\bibfield  {title} {\bibinfo {title} {The density-matrix renormalization group in the age of matrix product states},\ }\href {https://doi.org/https://doi.org/10.1016/j.aop.2010.09.012} {\bibfield  {journal} {\bibinfo  {journal} {Annals of Physics}\ }\textbf {\bibinfo {volume} {326}},\ \bibinfo {pages} {96} (\bibinfo {year} {2011})},\ \bibinfo {note} {{J}anuary 2011 Special Issue}\BibitemShut {NoStop}%
\bibitem [{\citenamefont {Nersesyan}(2020)}]{nersesyan_ssh_ladder}%
  \BibitemOpen
  \bibfield  {author} {\bibinfo {author} {\bibfnamefont {A.~A.}\ \bibnamefont {Nersesyan}},\ }\bibfield  {title} {\bibinfo {title} {Phase diagram of an interacting staggered {S}u-{S}chrieffer-{H}eeger two-chain ladder close to a quantum critical point},\ }\href {https://doi.org/10.1103/PhysRevB.102.045108} {\bibfield  {journal} {\bibinfo  {journal} {Phys. Rev. B}\ }\textbf {\bibinfo {volume} {102}},\ \bibinfo {pages} {045108} (\bibinfo {year} {2020})}\BibitemShut {NoStop}%
\bibitem [{\citenamefont {Padavi\ifmmode~\acute{c}\else \'{c}\fi{}}\ \emph {et~al.}(2018)\citenamefont {Padavi\ifmmode~\acute{c}\else \'{c}\fi{}}, \citenamefont {Hegde}, \citenamefont {DeGottardi},\ and\ \citenamefont {Vishveshwara}}]{smitha_ssh_ladder}%
  \BibitemOpen
  \bibfield  {author} {\bibinfo {author} {\bibfnamefont {K.}~\bibnamefont {Padavi\ifmmode~\acute{c}\else \'{c}\fi{}}}, \bibinfo {author} {\bibfnamefont {S.~S.}\ \bibnamefont {Hegde}}, \bibinfo {author} {\bibfnamefont {W.}~\bibnamefont {DeGottardi}},\ and\ \bibinfo {author} {\bibfnamefont {S.}~\bibnamefont {Vishveshwara}},\ }\bibfield  {title} {\bibinfo {title} {Topological phases, edge modes, and the {H}ofstadter butterfly in coupled {S}u-{S}chrieffer-{H}eeger systems},\ }\href {https://doi.org/10.1103/PhysRevB.98.024205} {\bibfield  {journal} {\bibinfo  {journal} {Phys. Rev. B}\ }\textbf {\bibinfo {volume} {98}},\ \bibinfo {pages} {024205} (\bibinfo {year} {2018})}\BibitemShut {NoStop}%
\bibitem [{\citenamefont {Padhan}\ \emph {et~al.}(2024)\citenamefont {Padhan}, \citenamefont {Mondal}, \citenamefont {Vishveshwara},\ and\ \citenamefont {Mishra}}]{padhan_ladder}%
  \BibitemOpen
  \bibfield  {author} {\bibinfo {author} {\bibfnamefont {A.}~\bibnamefont {Padhan}}, \bibinfo {author} {\bibfnamefont {S.}~\bibnamefont {Mondal}}, \bibinfo {author} {\bibfnamefont {S.}~\bibnamefont {Vishveshwara}},\ and\ \bibinfo {author} {\bibfnamefont {T.}~\bibnamefont {Mishra}},\ }\bibfield  {title} {\bibinfo {title} {Interacting bosons on a {S}u-{S}chrieffer-{S}eeger ladder: Topological phases and {T}houless pumping},\ }\href {https://doi.org/10.1103/PhysRevB.109.085120} {\bibfield  {journal} {\bibinfo  {journal} {Phys. Rev. B}\ }\textbf {\bibinfo {volume} {109}},\ \bibinfo {pages} {085120} (\bibinfo {year} {2024})}\BibitemShut {NoStop}%
\bibitem [{\citenamefont {Parida}\ \emph {et~al.}(2024)\citenamefont {Parida}, \citenamefont {Padhan},\ and\ \citenamefont {Mishra}}]{rajashri_ladder}%
  \BibitemOpen
  \bibfield  {author} {\bibinfo {author} {\bibfnamefont {R.}~\bibnamefont {Parida}}, \bibinfo {author} {\bibfnamefont {A.}~\bibnamefont {Padhan}},\ and\ \bibinfo {author} {\bibfnamefont {T.}~\bibnamefont {Mishra}},\ }\bibfield  {title} {\bibinfo {title} {Interaction driven topological phase transitions of hardcore bosons on a two-leg ladder},\ }\href {https://doi.org/10.1103/PhysRevB.110.165110} {\bibfield  {journal} {\bibinfo  {journal} {Phys. Rev. B}\ }\textbf {\bibinfo {volume} {110}},\ \bibinfo {pages} {165110} (\bibinfo {year} {2024})}\BibitemShut {NoStop}%
\bibitem [{\citenamefont {Block}\ \emph {et~al.}(2011)\citenamefont {Block}, \citenamefont {Mishmash}, \citenamefont {Kaul}, \citenamefont {Sheng}, \citenamefont {Motrunich},\ and\ \citenamefont {Fisher}}]{rmi_motrunich}%
  \BibitemOpen
  \bibfield  {author} {\bibinfo {author} {\bibfnamefont {M.~S.}\ \bibnamefont {Block}}, \bibinfo {author} {\bibfnamefont {R.~V.}\ \bibnamefont {Mishmash}}, \bibinfo {author} {\bibfnamefont {R.~K.}\ \bibnamefont {Kaul}}, \bibinfo {author} {\bibfnamefont {D.~N.}\ \bibnamefont {Sheng}}, \bibinfo {author} {\bibfnamefont {O.~I.}\ \bibnamefont {Motrunich}},\ and\ \bibinfo {author} {\bibfnamefont {M.~P.~A.}\ \bibnamefont {Fisher}},\ }\bibfield  {title} {\bibinfo {title} {Exotic gapless {M}ott insulators of bosons on multileg ladders},\ }\href {https://doi.org/10.1103/PhysRevLett.106.046402} {\bibfield  {journal} {\bibinfo  {journal} {Phys. Rev. Lett.}\ }\textbf {\bibinfo {volume} {106}},\ \bibinfo {pages} {046402} (\bibinfo {year} {2011})}\BibitemShut {NoStop}%
\bibitem [{\citenamefont {Carrasquilla}\ \emph {et~al.}(2011)\citenamefont {Carrasquilla}, \citenamefont {Becca},\ and\ \citenamefont {Fabrizio}}]{rmi_fabrizio}%
  \BibitemOpen
  \bibfield  {author} {\bibinfo {author} {\bibfnamefont {J.}~\bibnamefont {Carrasquilla}}, \bibinfo {author} {\bibfnamefont {F.}~\bibnamefont {Becca}},\ and\ \bibinfo {author} {\bibfnamefont {M.}~\bibnamefont {Fabrizio}},\ }\bibfield  {title} {\bibinfo {title} {Bose-glass, superfluid, and rung-{M}ott phases of hard-core bosons in disordered two-leg ladders},\ }\href {https://doi.org/10.1103/PhysRevB.83.245101} {\bibfield  {journal} {\bibinfo  {journal} {Phys. Rev. B}\ }\textbf {\bibinfo {volume} {83}},\ \bibinfo {pages} {245101} (\bibinfo {year} {2011})}\BibitemShut {NoStop}%
\bibitem [{\citenamefont {Cr\'epin}\ \emph {et~al.}(2011)\citenamefont {Cr\'epin}, \citenamefont {Laflorencie}, \citenamefont {Roux},\ and\ \citenamefont {Simon}}]{rmi_laflorencie}%
  \BibitemOpen
  \bibfield  {author} {\bibinfo {author} {\bibfnamefont {F.}~\bibnamefont {Cr\'epin}}, \bibinfo {author} {\bibfnamefont {N.}~\bibnamefont {Laflorencie}}, \bibinfo {author} {\bibfnamefont {G.}~\bibnamefont {Roux}},\ and\ \bibinfo {author} {\bibfnamefont {P.}~\bibnamefont {Simon}},\ }\bibfield  {title} {\bibinfo {title} {Phase diagram of hard-core bosons on clean and disordered two-leg ladders: {M}ott insulator--{L}uttinger liquid--{B}ose glass},\ }\href {https://doi.org/10.1103/PhysRevB.84.054517} {\bibfield  {journal} {\bibinfo  {journal} {Phys. Rev. B}\ }\textbf {\bibinfo {volume} {84}},\ \bibinfo {pages} {054517} (\bibinfo {year} {2011})}\BibitemShut {NoStop}%
\bibitem [{\citenamefont {Parida}\ \emph {et~al.}(2025{\natexlab{a}})\citenamefont {Parida}, \citenamefont {Paul}, \citenamefont {Padhi},\ and\ \citenamefont {Mishra}}]{rbi_parida_coupling}%
  \BibitemOpen
  \bibfield  {author} {\bibinfo {author} {\bibfnamefont {R.}~\bibnamefont {Parida}}, \bibinfo {author} {\bibfnamefont {B.}~\bibnamefont {Paul}}, \bibinfo {author} {\bibfnamefont {S.~R.}\ \bibnamefont {Padhi}},\ and\ \bibinfo {author} {\bibfnamefont {T.}~\bibnamefont {Mishra}},\ }\href@noop {} {\bibinfo {title} {Coupling induced emergent topology in a two-leg fermionic ladder}} (\bibinfo {year} {2025}{\natexlab{a}}),\ \Eprint {https://arxiv.org/abs/2509.25130} {arXiv:2509.25130 [cond-mat.quant-gas]} \BibitemShut {NoStop}%
\bibitem [{\citenamefont {Parida}\ \emph {et~al.}(2025{\natexlab{b}})\citenamefont {Parida}, \citenamefont {Sen},\ and\ \citenamefont {Mishra}}]{rajashri_four_quadrant}%
  \BibitemOpen
  \bibfield  {author} {\bibinfo {author} {\bibfnamefont {R.}~\bibnamefont {Parida}}, \bibinfo {author} {\bibfnamefont {D.}~\bibnamefont {Sen}},\ and\ \bibinfo {author} {\bibfnamefont {T.}~\bibnamefont {Mishra}},\ }\bibfield  {title} {\bibinfo {title} {Topological phase transition through tunable nearest-neighbor interactions in a one-dimensional lattice},\ }\href {https://doi.org/10.1103/kxqp-57vj} {\bibfield  {journal} {\bibinfo  {journal} {Phys. Rev. B}\ ,\ \bibinfo {pages} {085124}} (\bibinfo {year} {2025}{\natexlab{b}})}\BibitemShut {NoStop}%
\bibitem [{\citenamefont {Thouless}(1983)}]{Thouless1983}%
  \BibitemOpen
  \bibfield  {author} {\bibinfo {author} {\bibfnamefont {D.~J.}\ \bibnamefont {Thouless}},\ }\bibfield  {title} {\bibinfo {title} {Quantization of particle transport},\ }\href {https://doi.org/10.1103/PhysRevB.27.6083} {\bibfield  {journal} {\bibinfo  {journal} {Phys. Rev. B}\ }\textbf {\bibinfo {volume} {27}},\ \bibinfo {pages} {6083} (\bibinfo {year} {1983})}\BibitemShut {NoStop}%
\bibitem [{\citenamefont {Citro}\ and\ \citenamefont {Aidelsburger}(2023)}]{monika_review}%
  \BibitemOpen
  \bibfield  {author} {\bibinfo {author} {\bibfnamefont {R.}~\bibnamefont {Citro}}\ and\ \bibinfo {author} {\bibfnamefont {M.}~\bibnamefont {Aidelsburger}},\ }\bibfield  {title} {\bibinfo {title} {{T}houless pumping and topology},\ }\href {https://doi.org/10.1038/s42254-022-00545-0} {\bibfield  {journal} {\bibinfo  {journal} {Nature Reviews Physics}\ }\textbf {\bibinfo {volume} {5}},\ \bibinfo {pages} {87} (\bibinfo {year} {2023})}\BibitemShut {NoStop}%
\bibitem [{\citenamefont {Lin}\ \emph {et~al.}(2020)\citenamefont {Lin}, \citenamefont {Ke},\ and\ \citenamefont {Lee}}]{bound_pump1}%
  \BibitemOpen
  \bibfield  {author} {\bibinfo {author} {\bibfnamefont {L.}~\bibnamefont {Lin}}, \bibinfo {author} {\bibfnamefont {Y.}~\bibnamefont {Ke}},\ and\ \bibinfo {author} {\bibfnamefont {C.}~\bibnamefont {Lee}},\ }\bibfield  {title} {\bibinfo {title} {Interaction-induced topological bound states and {T}houless pumping in a one-dimensional optical lattice},\ }\href {https://doi.org/10.1103/PhysRevA.101.023620} {\bibfield  {journal} {\bibinfo  {journal} {Phys. Rev. A}\ }\textbf {\bibinfo {volume} {101}},\ \bibinfo {pages} {023620} (\bibinfo {year} {2020})}\BibitemShut {NoStop}%
\bibitem [{\citenamefont {Ke}\ \emph {et~al.}(2017)\citenamefont {Ke}, \citenamefont {Qin}, \citenamefont {Kivshar},\ and\ \citenamefont {Lee}}]{bound_pump3}%
  \BibitemOpen
  \bibfield  {author} {\bibinfo {author} {\bibfnamefont {Y.}~\bibnamefont {Ke}}, \bibinfo {author} {\bibfnamefont {X.}~\bibnamefont {Qin}}, \bibinfo {author} {\bibfnamefont {Y.~S.}\ \bibnamefont {Kivshar}},\ and\ \bibinfo {author} {\bibfnamefont {C.}~\bibnamefont {Lee}},\ }\bibfield  {title} {\bibinfo {title} {Multiparticle {W}annier states and {T}houless pumping of interacting bosons},\ }\href {https://doi.org/10.1103/PhysRevA.95.063630} {\bibfield  {journal} {\bibinfo  {journal} {Phys. Rev. A}\ }\textbf {\bibinfo {volume} {95}},\ \bibinfo {pages} {063630} (\bibinfo {year} {2017})}\BibitemShut {NoStop}%
\bibitem [{\citenamefont {Kuno}\ \emph {et~al.}(2017)\citenamefont {Kuno}, \citenamefont {Shimizu},\ and\ \citenamefont {Ichinose}}]{Kuno2017}%
  \BibitemOpen
  \bibfield  {author} {\bibinfo {author} {\bibfnamefont {Y.}~\bibnamefont {Kuno}}, \bibinfo {author} {\bibfnamefont {K.}~\bibnamefont {Shimizu}},\ and\ \bibinfo {author} {\bibfnamefont {I.}~\bibnamefont {Ichinose}},\ }\bibfield  {title} {\bibinfo {title} {Various topological {M}ott insulators and topological bulk charge pumping in strongly-interacting boson system in one-dimensional superlattice},\ }\href@noop {} {\bibfield  {journal} {\bibinfo  {journal} {New Journal of Physics}\ }\textbf {\bibinfo {volume} {19}},\ \bibinfo {pages} {123025} (\bibinfo {year} {2017})}\BibitemShut {NoStop}%
\bibitem [{\citenamefont {Bertok}\ \emph {et~al.}(2022)\citenamefont {Bertok}, \citenamefont {Heidrich-Meisner},\ and\ \citenamefont {Aligia}}]{bertok_pump}%
  \BibitemOpen
  \bibfield  {author} {\bibinfo {author} {\bibfnamefont {E.}~\bibnamefont {Bertok}}, \bibinfo {author} {\bibfnamefont {F.}~\bibnamefont {Heidrich-Meisner}},\ and\ \bibinfo {author} {\bibfnamefont {A.~A.}\ \bibnamefont {Aligia}},\ }\bibfield  {title} {\bibinfo {title} {Splitting of topological charge pumping in an interacting two-component fermionic {R}ice-{M}ele {H}ubbard model},\ }\href {https://doi.org/10.1103/PhysRevB.106.045141} {\bibfield  {journal} {\bibinfo  {journal} {Phys. Rev. B}\ }\textbf {\bibinfo {volume} {106}},\ \bibinfo {pages} {045141} (\bibinfo {year} {2022})}\BibitemShut {NoStop}%
\bibitem [{\citenamefont {Mondal}\ \emph {et~al.}(2022)\citenamefont {Mondal}, \citenamefont {Bertok},\ and\ \citenamefont {Heidrich-Meisner}}]{mondal_phonon}%
  \BibitemOpen
  \bibfield  {author} {\bibinfo {author} {\bibfnamefont {S.}~\bibnamefont {Mondal}}, \bibinfo {author} {\bibfnamefont {E.}~\bibnamefont {Bertok}},\ and\ \bibinfo {author} {\bibfnamefont {F.}~\bibnamefont {Heidrich-Meisner}},\ }\bibfield  {title} {\bibinfo {title} {Phonon-induced breakdown of {T}houless pumping in the {R}ice-{M}ele-{H}olstein model},\ }\href {https://doi.org/10.1103/PhysRevB.106.235118} {\bibfield  {journal} {\bibinfo  {journal} {Phys. Rev. B}\ }\textbf {\bibinfo {volume} {106}},\ \bibinfo {pages} {235118} (\bibinfo {year} {2022})}\BibitemShut {NoStop}%
\bibitem [{\citenamefont {Hayward}\ \emph {et~al.}(2018)\citenamefont {Hayward}, \citenamefont {Schweizer}, \citenamefont {Lohse}, \citenamefont {Aidelsburger},\ and\ \citenamefont {Heidrich-Meisner}}]{Hayward2018}%
  \BibitemOpen
  \bibfield  {author} {\bibinfo {author} {\bibfnamefont {A.}~\bibnamefont {Hayward}}, \bibinfo {author} {\bibfnamefont {C.}~\bibnamefont {Schweizer}}, \bibinfo {author} {\bibfnamefont {M.}~\bibnamefont {Lohse}}, \bibinfo {author} {\bibfnamefont {M.}~\bibnamefont {Aidelsburger}},\ and\ \bibinfo {author} {\bibfnamefont {F.}~\bibnamefont {Heidrich-Meisner}},\ }\bibfield  {title} {\bibinfo {title} {Topological charge pumping in the interacting bosonic {R}ice-{M}ele model},\ }\href {https://doi.org/10.1103/PhysRevB.98.245148} {\bibfield  {journal} {\bibinfo  {journal} {Phys. Rev. B}\ }\textbf {\bibinfo {volume} {98}},\ \bibinfo {pages} {245148} (\bibinfo {year} {2018})}\BibitemShut {NoStop}%
\bibitem [{\citenamefont {Schweizer}\ \emph {et~al.}(2016)\citenamefont {Schweizer}, \citenamefont {Lohse}, \citenamefont {Citro},\ and\ \citenamefont {Bloch}}]{spin_pumping}%
  \BibitemOpen
  \bibfield  {author} {\bibinfo {author} {\bibfnamefont {C.}~\bibnamefont {Schweizer}}, \bibinfo {author} {\bibfnamefont {M.}~\bibnamefont {Lohse}}, \bibinfo {author} {\bibfnamefont {R.}~\bibnamefont {Citro}},\ and\ \bibinfo {author} {\bibfnamefont {I.}~\bibnamefont {Bloch}},\ }\bibfield  {title} {\bibinfo {title} {Spin pumping and measurement of spin currents in optical superlattices},\ }\href {https://doi.org/10.1103/PhysRevLett.117.170405} {\bibfield  {journal} {\bibinfo  {journal} {Phys. Rev. Lett.}\ }\textbf {\bibinfo {volume} {117}},\ \bibinfo {pages} {170405} (\bibinfo {year} {2016})}\BibitemShut {NoStop}%
\bibitem [{\citenamefont {Kraus}\ \emph {et~al.}(2012)\citenamefont {Kraus}, \citenamefont {Lahini}, \citenamefont {Ringel}, \citenamefont {Verbin},\ and\ \citenamefont {Zilberberg}}]{pumping_quasicrystals}%
  \BibitemOpen
  \bibfield  {author} {\bibinfo {author} {\bibfnamefont {Y.~E.}\ \bibnamefont {Kraus}}, \bibinfo {author} {\bibfnamefont {Y.}~\bibnamefont {Lahini}}, \bibinfo {author} {\bibfnamefont {Z.}~\bibnamefont {Ringel}}, \bibinfo {author} {\bibfnamefont {M.}~\bibnamefont {Verbin}},\ and\ \bibinfo {author} {\bibfnamefont {O.}~\bibnamefont {Zilberberg}},\ }\bibfield  {title} {\bibinfo {title} {Topological states and adiabatic pumping in quasicrystals},\ }\href {https://doi.org/10.1103/PhysRevLett.109.106402} {\bibfield  {journal} {\bibinfo  {journal} {Phys. Rev. Lett.}\ }\textbf {\bibinfo {volume} {109}},\ \bibinfo {pages} {106402} (\bibinfo {year} {2012})}\BibitemShut {NoStop}%
\bibitem [{\citenamefont {Wang}\ \emph {et~al.}(2013)\citenamefont {Wang}, \citenamefont {Troyer},\ and\ \citenamefont {Dai}}]{pumping_1d}%
  \BibitemOpen
  \bibfield  {author} {\bibinfo {author} {\bibfnamefont {L.}~\bibnamefont {Wang}}, \bibinfo {author} {\bibfnamefont {M.}~\bibnamefont {Troyer}},\ and\ \bibinfo {author} {\bibfnamefont {X.}~\bibnamefont {Dai}},\ }\bibfield  {title} {\bibinfo {title} {Topological charge pumping in a one-dimensional optical lattice},\ }\href {https://doi.org/10.1103/PhysRevLett.111.026802} {\bibfield  {journal} {\bibinfo  {journal} {Phys. Rev. Lett.}\ }\textbf {\bibinfo {volume} {111}},\ \bibinfo {pages} {026802} (\bibinfo {year} {2013})}\BibitemShut {NoStop}%
\bibitem [{\citenamefont {Walter}\ \emph {et~al.}(2023)\citenamefont {Walter}, \citenamefont {Zhu}, \citenamefont {Gächter}, \citenamefont {Minguzzi}, \citenamefont {Roschinski}, \citenamefont {Sandholzer}, \citenamefont {Viebahn},\ and\ \citenamefont {Esslinger}}]{hubbarad_thouless_pump}%
  \BibitemOpen
  \bibfield  {author} {\bibinfo {author} {\bibfnamefont {A.-S.}\ \bibnamefont {Walter}}, \bibinfo {author} {\bibfnamefont {Z.}~\bibnamefont {Zhu}}, \bibinfo {author} {\bibfnamefont {M.}~\bibnamefont {Gächter}}, \bibinfo {author} {\bibfnamefont {J.}~\bibnamefont {Minguzzi}}, \bibinfo {author} {\bibfnamefont {S.}~\bibnamefont {Roschinski}}, \bibinfo {author} {\bibfnamefont {K.}~\bibnamefont {Sandholzer}}, \bibinfo {author} {\bibfnamefont {K.}~\bibnamefont {Viebahn}},\ and\ \bibinfo {author} {\bibfnamefont {T.}~\bibnamefont {Esslinger}},\ }\bibfield  {title} {\bibinfo {title} {Quantization and its breakdown in a {H}ubbard–{T}houless pump},\ }\href {https://doi.org/10.1038/s41567-023-02145-w} {\bibfield  {journal} {\bibinfo  {journal} {Nature Physics}\ }\textbf {\bibinfo {volume} {19}},\ \bibinfo {pages} {1471–1475} (\bibinfo {year} {2023})}\BibitemShut {NoStop}%
\bibitem [{\citenamefont {Ke}\ \emph {et~al.}(2020)\citenamefont {Ke}, \citenamefont {Zhong}, \citenamefont {Poshakinskiy}, \citenamefont {Kivshar}, \citenamefont {Poddubny},\ and\ \citenamefont {Lee}}]{qubit_pumping}%
  \BibitemOpen
  \bibfield  {author} {\bibinfo {author} {\bibfnamefont {Y.}~\bibnamefont {Ke}}, \bibinfo {author} {\bibfnamefont {J.}~\bibnamefont {Zhong}}, \bibinfo {author} {\bibfnamefont {A.~V.}\ \bibnamefont {Poshakinskiy}}, \bibinfo {author} {\bibfnamefont {Y.~S.}\ \bibnamefont {Kivshar}}, \bibinfo {author} {\bibfnamefont {A.~N.}\ \bibnamefont {Poddubny}},\ and\ \bibinfo {author} {\bibfnamefont {C.}~\bibnamefont {Lee}},\ }\bibfield  {title} {\bibinfo {title} {Radiative topological biphoton states in modulated qubit arrays},\ }\href {https://doi.org/10.1103/PhysRevResearch.2.033190} {\bibfield  {journal} {\bibinfo  {journal} {Phys. Rev. Res.}\ }\textbf {\bibinfo {volume} {2}},\ \bibinfo {pages} {033190} (\bibinfo {year} {2020})}\BibitemShut {NoStop}%
\bibitem [{\citenamefont {J{\"u}rgensen}\ \emph {et~al.}(2023)\citenamefont {J{\"u}rgensen}, \citenamefont {Mukherjee}, \citenamefont {J{\"o}rg},\ and\ \citenamefont {Rechtsman}}]{seba_nphy_pumming}%
  \BibitemOpen
  \bibfield  {author} {\bibinfo {author} {\bibfnamefont {M.}~\bibnamefont {J{\"u}rgensen}}, \bibinfo {author} {\bibfnamefont {S.}~\bibnamefont {Mukherjee}}, \bibinfo {author} {\bibfnamefont {C.}~\bibnamefont {J{\"o}rg}},\ and\ \bibinfo {author} {\bibfnamefont {M.~C.}\ \bibnamefont {Rechtsman}},\ }\bibfield  {title} {\bibinfo {title} {Quantized fractional {T}houless pumping of solitons},\ }\href {https://doi.org/10.1038/s41567-022-01871-x} {\bibfield  {journal} {\bibinfo  {journal} {Nature Physics}\ }\textbf {\bibinfo {volume} {19}},\ \bibinfo {pages} {420} (\bibinfo {year} {2023})}\BibitemShut {NoStop}%
\bibitem [{\citenamefont {J{\"u}rgensen}\ \emph {et~al.}(2021)\citenamefont {J{\"u}rgensen}, \citenamefont {Mukherjee},\ and\ \citenamefont {Rechtsman}}]{seba_nature_pumping}%
  \BibitemOpen
  \bibfield  {author} {\bibinfo {author} {\bibfnamefont {M.}~\bibnamefont {J{\"u}rgensen}}, \bibinfo {author} {\bibfnamefont {S.}~\bibnamefont {Mukherjee}},\ and\ \bibinfo {author} {\bibfnamefont {M.~C.}\ \bibnamefont {Rechtsman}},\ }\bibfield  {title} {\bibinfo {title} {Quantized nonlinear {T}houless pumping},\ }\href {https://doi.org/10.1038/s41586-021-03688-9} {\bibfield  {journal} {\bibinfo  {journal} {Nature}\ }\textbf {\bibinfo {volume} {596}},\ \bibinfo {pages} {63} (\bibinfo {year} {2021})}\BibitemShut {NoStop}%
\bibitem [{\citenamefont {Viebahn}\ \emph {et~al.}(2024)\citenamefont {Viebahn}, \citenamefont {Walter}, \citenamefont {Bertok}, \citenamefont {Zhu}, \citenamefont {G\"achter}, \citenamefont {Aligia}, \citenamefont {Heidrich-Meisner},\ and\ \citenamefont {Esslinger}}]{pumping_esslinger}%
  \BibitemOpen
  \bibfield  {author} {\bibinfo {author} {\bibfnamefont {K.}~\bibnamefont {Viebahn}}, \bibinfo {author} {\bibfnamefont {A.-S.}\ \bibnamefont {Walter}}, \bibinfo {author} {\bibfnamefont {E.}~\bibnamefont {Bertok}}, \bibinfo {author} {\bibfnamefont {Z.}~\bibnamefont {Zhu}}, \bibinfo {author} {\bibfnamefont {M.}~\bibnamefont {G\"achter}}, \bibinfo {author} {\bibfnamefont {A.~A.}\ \bibnamefont {Aligia}}, \bibinfo {author} {\bibfnamefont {F.}~\bibnamefont {Heidrich-Meisner}},\ and\ \bibinfo {author} {\bibfnamefont {T.}~\bibnamefont {Esslinger}},\ }\bibfield  {title} {\bibinfo {title} {Interactions enable {T}houless pumping in a nonsliding lattice},\ }\href {https://doi.org/10.1103/PhysRevX.14.021049} {\bibfield  {journal} {\bibinfo  {journal} {Phys. Rev. X}\ }\textbf {\bibinfo {volume} {14}},\ \bibinfo {pages} {021049} (\bibinfo {year} {2024})}\BibitemShut {NoStop}%
\bibitem [{\citenamefont {Zhu}\ \emph {et~al.}(2024)\citenamefont {Zhu}, \citenamefont {Gächter}, \citenamefont {Walter}, \citenamefont {Viebahn},\ and\ \citenamefont {Esslinger}}]{pump_reversal_esslinger}%
  \BibitemOpen
  \bibfield  {author} {\bibinfo {author} {\bibfnamefont {Z.}~\bibnamefont {Zhu}}, \bibinfo {author} {\bibfnamefont {M.}~\bibnamefont {Gächter}}, \bibinfo {author} {\bibfnamefont {A.-S.}\ \bibnamefont {Walter}}, \bibinfo {author} {\bibfnamefont {K.}~\bibnamefont {Viebahn}},\ and\ \bibinfo {author} {\bibfnamefont {T.}~\bibnamefont {Esslinger}},\ }\bibfield  {title} {\bibinfo {title} {Reversal of quantized hall drifts at noninteracting and interacting topological boundaries},\ }\href {https://doi.org/10.1126/science.adg3848} {\bibfield  {journal} {\bibinfo  {journal} {Science}\ }\textbf {\bibinfo {volume} {384}},\ \bibinfo {pages} {317} (\bibinfo {year} {2024})}\BibitemShut {NoStop}%
\bibitem [{\citenamefont {Rice}\ and\ \citenamefont {Mele}(1982)}]{rice_mele_model}%
  \BibitemOpen
  \bibfield  {author} {\bibinfo {author} {\bibfnamefont {M.~J.}\ \bibnamefont {Rice}}\ and\ \bibinfo {author} {\bibfnamefont {E.~J.}\ \bibnamefont {Mele}},\ }\bibfield  {title} {\bibinfo {title} {Elementary excitations of a linearly conjugated diatomic polymer},\ }\href {https://doi.org/10.1103/PhysRevLett.49.1455} {\bibfield  {journal} {\bibinfo  {journal} {Phys. Rev. Lett.}\ }\textbf {\bibinfo {volume} {49}},\ \bibinfo {pages} {1455} (\bibinfo {year} {1982})}\BibitemShut {NoStop}%
\bibitem [{\citenamefont {Berry}(1984)}]{berry_phase}%
  \BibitemOpen
  \bibfield  {author} {\bibinfo {author} {\bibfnamefont {M.~V.}\ \bibnamefont {Berry}},\ }\bibfield  {title} {\bibinfo {title} {Quantal phase factors accompanying adiabatic changes},\ }\href {https://doi.org/10.1098/rspa.1984.0023} {\bibfield  {journal} {\bibinfo  {journal} {Proc. Roy. Soc. London}\ }\textbf {\bibinfo {volume} {A 392}},\ \bibinfo {pages} {45} (\bibinfo {year} {1984})}\BibitemShut {NoStop}%
\bibitem [{\citenamefont {Mao}\ \emph {et~al.}(2026)\citenamefont {Mao}, \citenamefont {Zeng}, \citenamefont {Wu}, \citenamefont {Xie}, \citenamefont {Yuan}, \citenamefont {Chen}, \citenamefont {Dai},\ and\ \citenamefont {Pan}}]{expt_ssh_ladder}%
  \BibitemOpen
  \bibfield  {author} {\bibinfo {author} {\bibfnamefont {Y.-Y.}\ \bibnamefont {Mao}}, \bibinfo {author} {\bibfnamefont {C.}~\bibnamefont {Zeng}}, \bibinfo {author} {\bibfnamefont {F.-F.}\ \bibnamefont {Wu}}, \bibinfo {author} {\bibfnamefont {Y.-J.}\ \bibnamefont {Xie}}, \bibinfo {author} {\bibfnamefont {T.}~\bibnamefont {Yuan}}, \bibinfo {author} {\bibfnamefont {Y.-A.}\ \bibnamefont {Chen}}, \bibinfo {author} {\bibfnamefont {H.-N.}\ \bibnamefont {Dai}},\ and\ \bibinfo {author} {\bibfnamefont {J.-W.}\ \bibnamefont {Pan}},\ }\bibfield  {title} {\bibinfo {title} {Realizing a topological staggered {S}u-{S}chrieffer-{H}eeger ladder with atomic momentum lattice},\ }\href {https://doi.org/10.1103/3r1y-81lc} {\bibfield  {journal} {\bibinfo  {journal} {Phys. Rev. Res.}\ }\textbf {\bibinfo {volume} {8}},\ \bibinfo {pages} {L022006} (\bibinfo {year} {2026})}\BibitemShut {NoStop}%
\end{thebibliography}%

\clearpage

\vspace*{.4cm}

\appendix

\onecolumngrid

\begin{center}
\bf{END MATTER}
\end{center}


\twocolumngrid

\subsection{Hartree-Fock Approximation in the limit of $t_1 = t_p 
= 1$ and $t_2 \ll 1$}

We present here the details of the Hartree-Fock approximation mentioned in the main text.
We consider a two-leg ladder model where the upper leg has hoppings 
$-t_1, -t_2, -t_1,
-t_2, \cdots$, and the lower leg has hoppings $-t_2, -t_1, -t_2, -t_1, 
\cdots$. There is also a hopping $-t_p$ between neighboring sites on the
upper and lower legs. We can write all this in a compact way in a
spinless fermionic Hamiltonian as
\beq H_f ~=~ - \sum_{ij} ~t_{ij} (c_i^\da c_j + c_j^\da c_i), 
\label{hamf} \eeq
and in a spin-1/2 Hamiltonian as
\beq H_s ~=~ - \sum_{ij} ~t_{ij} (\si_i^+ \si_j^- + \si_j^+ \si_j^-), 
\label{hams} \eeq
with the convention that at each site $i$, $\si_i^z = 1$ denotes a 
single hard-core boson and $\si_i^z = -1$ denotes an empty site.

We first comment on a symmetry satisfied by the properties of an infinitely
large system described by either of the two Hamiltonians given above.
Under an exchange of the upper and lower legs, we have $t_1 \leftrightarrow 
t_2$. Setting $(t_1,t_2,t_p) = (1,t_2,t_p)$, we can rescale it to 
$(1/t_2,1,t_p/t_2)$ (since an overall scaling of the 
Hamiltonian by $t_2$ does not change the phase of the system) and then use 
the exchange symmetry to transform this to $(1,1/t_2,t_p/t_2)$. Hence the
system must be in the same phase at the points $(1,t_2,t_p)$ and 
$(1,1/t_2,t_p/t_2)$.

We now consider the fermionic system described in Eq.~\eqref{hamf}. We
find that in the phase diagram in the two-dimensional space described by
$(t_p,t_2)$ (setting $t_1=1$), there are two gapless lines, one given by 
$t_2 = 1$ and $0 \le t_p \le 2$ and another given by $t_p = t_1 + t_2
= 1 + t_2$.

Next, we consider the spin-1/2 (hard-core boson) model described in 
Eq.~\eqref{hams}. We find numerically that there are two gapless lines,
one going from $(t_p,t_2)=(1,0)$ to $(0,1)$ and the other going from
$(1,0)$ to $(\infty,\infty)$ along some direction. These two lines are
related by the exchange symmetry mentioned above. It is therefore sufficient
to study the lower line going from $(1,0)$ to $(0,1)$. To understand this
line, we first look at the neighborhood of the point $(1,0)$. For $t_1 = t_p 
= 1$ and $t_2 \ll 1$, it is useful to define a Jordan-Wigner (JW) 
transformation from spins to fermions. We take the JW string of $\si_i^z$'s to 
go along the bonds with hoppings $t_1, t_p, t_1, t_p, \cdots$. Numbering these 
sites by an integer $n$ which goes from $- \infty$ to $\infty$, we have the JW 
transformation
\bea d_n &=& \left( \prod_{k=-\infty}^{n-1} ~(- \si^z_k) \right) ~\si_n^-, 
\non \\
d_n^\da &=& \left( \prod_{k=-\infty}^{n-1} ~(- \si^z_k) \right) ~\si_n^+. 
\label{jw} \eea
We note that 
\beq \si_n^z ~=~ 2 d_n^\da d_n -1. \label{sz} \eeq
Eq.~\eqref{jw} exactly transforms the $t_1, t_p$ bonds from the form given in 
Eq.~\eqref{hams} to the form given in Eq.~\eqref{hamf}. However, the $t_2$ 
bonds look different from what we see in Eq.~\eqref{hamf}. Using
Eqs.~\eqref{jw} and \eqref{sz}, we find that
terms like $-t_2 (\si_n^+ \si_{n+3}^- + \si_n^+ \si_{n+3}^-)$ in
Eq.~\eqref{hams} transform to $-t_2 (d_n^\da d_{n+3} + d_{n+3}^\da d_n) 
(2 d_{n+1}^\da d_{n+1} - 1)(2 d_{n+2}^\da d_{n+2} - 1)$. This is an 
interaction term since it involves up to six fermion operators. Assuming that
$t_2$ is small, we will now treat this term perturbatively using a 
Hartree-Fock approximation. 
Namely, we replace $(2 d_{n+1}^\da d_{n+1} - 1)(2 d_{n+2}^\da d_{n+2} - 1)$ 
by its expectation value in the ground state of the system with $t_1 = t_p =1$,
$t_2 = 0$ and an arbitrary filling $\rho$. Going to momentum space where
the momentum $k$ lies in the range $- \pi$ to $\pi$, the ground state has
all the states from $-k_F$ to $k_F$ filled and the remaining states empty, 
where $\rho = k_F /\pi$. Further,
\bea d_n &=& \int_{-\pi}^\pi \frac{dk}{2\pi} ~d_k e^{ikn}, \non \\  
d_n^\da &=& \int_{-\pi}^\pi \frac{dk}{2\pi} ~d_k^\da e^{-ikn}, \eea
where 
\bea \{ d_k, d_{k'} \} &=& \{ d_k^\da, d_{k'}^\da \} = 0, \non \\
\{ d_k, d_{k'}^\da \} &=& 2 \pi \de (k-k'). \label{anticomm} \eea
We then have
\begin{align} &\la (2 d_{n+1}^\da d_{n+1} - 1)(2 d_{n+2}^\da d_{n+2} - 1) \ra \non\\  & =
 4 \la d_{n+2}^\da d_{n+1}^\da d_{n+1} d_{n+2} \ra ~-~ 4 \la d_{n+1}^\da d_{n+1}
\ra + 1. \label{hf1} \end{align}
\begin{widetext}
The first term on the RHS of Eq.~\eqref{hf1} is given by
\bea && 4 \int \frac{dk_1}{2 \pi} ~\int \frac{dk_2}{2 \pi} \int 
\frac{dk_3}{2 \pi} \int \frac{dk_4}{2 \pi} ~\la c_{k_4}^\da 
c_{k_3}^\da c_{k_2} c_{k_1} \ra ~e^{-ik_4 (n+2) - ik_3 (n+1) + ik_2 (n+1) + 
ik_1 (n+2)} \non \\
&=& 4 \int_{-k_F}^{k_F} \frac{dk_1}{2 \pi} ~\int_{-k_F}^{k_F} \frac{dk_2}{2 
\pi} \int \frac{dk_3}{2 \pi} \int \frac{dk_4}{2 \pi} ~
[(2\pi)^2 \de (k_1 - k_4) \de (k_2 - k_3)
- (2\pi)^2 \de (k_1 - k_3) \de (k_2 - k_4) e^{i (k_1 - k_2)}] \non \\
&=& 4 \int_{-k_F}^{k_F} \frac{dk_1}{2 \pi} ~\int_{-k_F}^{k_F} \frac{dk_2}{2
\pi} ~[ 1 - e^{i (k_1 - k_2)}] \non \\
&=& \frac{4~}{\pi^2} \left[ k_F^2 - \sin^2 k_F \right]. \label{first} \eea
\end{widetext}
The second term on the RHS of Eq.~\eqref{hf1} is given by $- 4 k_F/\pi$.
Putting everything together, we obtain
\bea && \la (2 d_{n+1}^\da d_{n+1} - 1)(2 d_{n+2}^\da d_{n+2} - 1) \ra \non \\
&& =~ \frac{4~}{\pi^2} \left[ k_F^2 - \sin^2 k_F \right] ~-~ 
\frac{4k_F}{\pi} ~+~ 1. \label{hf2} \eea
For half-filling, $k_F = \pi/2$, we obtain
\beq \la (2 d_{n+1}^\da d_{n+1} - 1)(2 d_{n+2}^\da d_{n+2} - 1) \ra ~=~ - 
\frac{4~}{\pi^2}. \label{hf3} \eeq
We therefore see that the Hartree-Fock approximation effectively changes
the coefficient of the hopping amplitude from $t_2$ to $- (4/\pi^2) t_2$
in the Hamiltonian for the $d$-fermions.

Now, since the gapless line for the $c$-fermions (Eq.~\eqref{hamf}) is
known to be $t_p = 1 + t_2$ (setting $t_1 = 1$), we conclude that the gapless
line for the $d$-fermions should be given by 
\beq t_p ~=~ 1 ~-~ \frac{4~}{\pi^2} ~t_2 \eeq
for small $t_2$. Hence the slope $dt_2 /dt_p$ of the line at the point 
$(t_p = 1, t_2 = 0)$ should be equal to $- \pi^2 /4 \simeq - 2.47$. This qualitatively agrees with the
slope of the line that emerges from that point
in Fig.~\ref{fig:phase_diagram}(b) in the main text.

\begin{figure}[b]
\begin{center}
\includegraphics[width=\columnwidth]{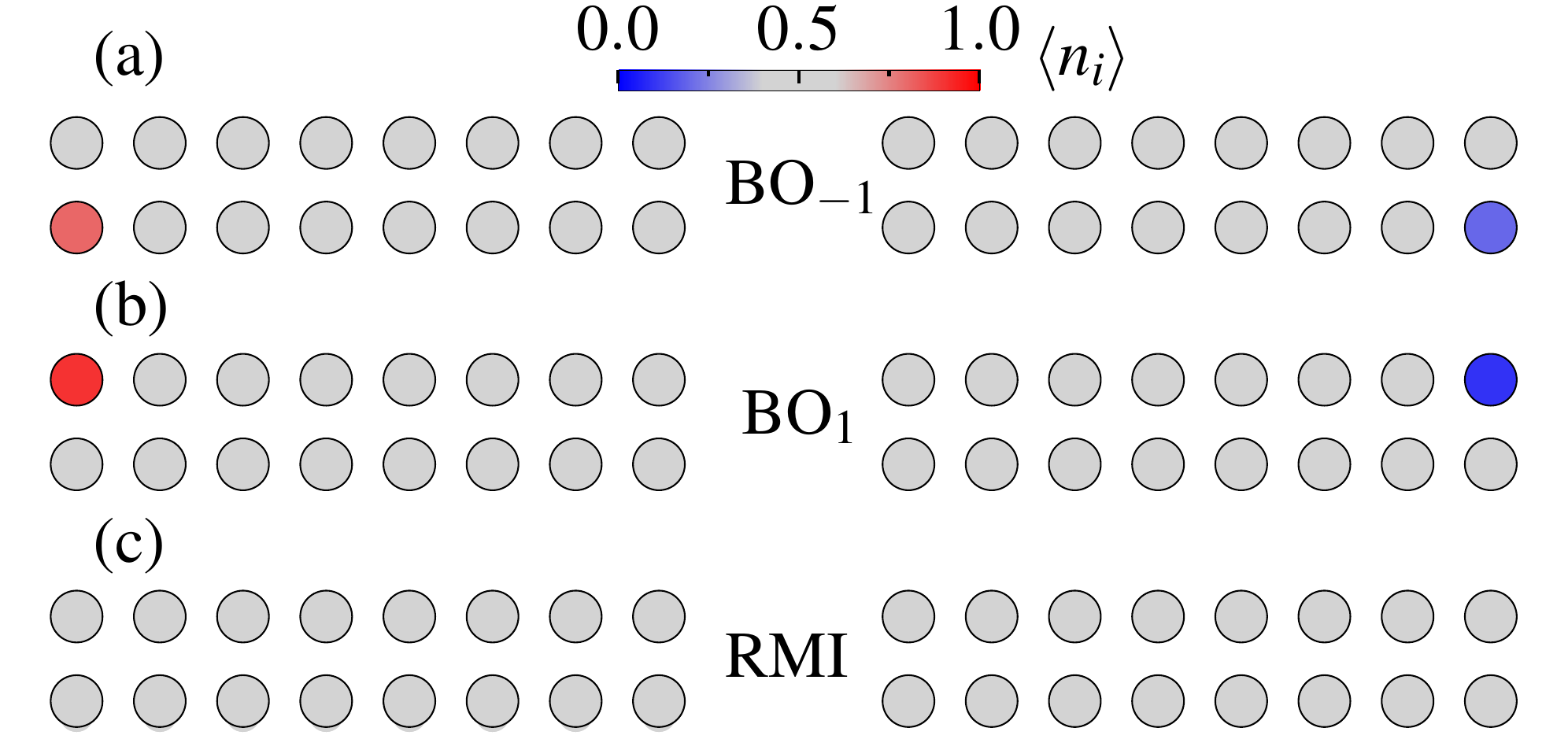}
\end{center}
\caption{The on-site particle density ($\langle n_i \rangle$)  for different phases corresponding to Fig.~\ref{fig:phase_diagram}(b) with a system consisting of $L = 100$ rungs. In the figures, the face color of the circles represents the values of $\langle n_i \rangle$ at $1/2$ filling ($N = L$) for (a) $t_p=0.2$, $t_2 = 0.3$, (b)  $t_p=0.2$, $t_2 = 1.8$, (c) $t_p=2.0$, $t_2 = 0.3$. For all the plots, the value of $t_1=1.0$ and only the edge sites have been shown to indicate the edge states in the system.}
\label{fig:edge_state}
\end{figure}

\subsection{Appearance of edge states}
The appearance of topology as well as edge states, for the case of hardcore bosons, has already been explained rigorously in Ref.~\cite{padhan_ladder}. 
Here, we show the edge states for our model of
hardcore bosons for completeness. 
%
%
We compute the on-site particle density in the BO$_{-1}$ ($t_p=0.2,~t_2=0.3$), BO$_1$ ($t_p=2.0,~t_2=0.3$), and RMI ($t_p=1.8,~t_2=0.2$) phases corresponding to Fig.~\ref{fig:phase_diagram}(b) for a system of $L=100$ rungs under OBC, and 
present the results in Figs.~\ref{fig:edge_state}(a–c), respectively, displaying only a few edge rungs for clarity. 
In Fig.~\ref{fig:edge_state}, the facecolor of the circles represents the onsite particle density.
A clear edge density imbalance in panels (a) and (b) suggests a topological bulk, whereas the uniform density at $t_p=2.0,t_2=0.3$ [Fig.~\ref{fig:edge_state}(c)] signals towards a trivial phase.
A similar behavior is observed in the spinless fermionic case for suitably chosen parameters (not shown here).

\end{document}